\documentclass{aa}
\usepackage{txfonts,graphicx,natbib}

\def\xmm{{\sl XMM-Newton}}
\def\c{{\sl Chandra}}
\def\einstein{{\sl Einstein}}
\def\rosat{{\sl ROSAT}}

\def\ltsim{\mathrel{\hbox{\rlap{\hbox{\lower4pt\hbox{$\sim$}}}\hbox{$<$}}}}
\def\gtsim{\mathrel{\hbox{\rlap{\hbox{\lower4pt\hbox{$\sim$}}}\hbox{$>$}}}}
\def\ergpspsqcm{erg~s$^{-1}$~cm$^{-2}$}

\def\sqd{deg$^2$}

\def\nh{$N_{\rm H}$}

\def\lognlogs{log$N$--log$S$}
\def\hr{{\em HR}}
\font\manual=manfnt 
\def\tickmark{\manual\char'170\rm\kern.15em}

\def\oiii{[O{\sc iii}]}

\def\nii{[N{\sc ii}]}

\begin{document}

\title{The XMM Large Scale Structure Survey: Properties and Two-Point Angular Correlations of Point-like Sources
\thanks{Based on observations obtained with \xmm, an ESA science mission with instruments and contributions directly funded by ESA Member States and NASA.}
}

\author{P.~Gandhi\inst{1,2} \and O.~Garcet\inst{3} \and L.~Disseau\inst{4,2} \and F.~Pacaud\inst{5} \and M.~Pierre\inst{5} \and A.~Gueguen\inst{5} \and D.~Alloin\inst{5,2} \and L.~Chiappetti\inst{6} \and E.~Gosset\inst{3} \and D.~Maccagni\inst{6} \and J.~Surdej\inst{3} \and I. Valtchanov\inst{7}
       }
\institute{Institute of Astronomy, Madingley Road, University of Cambridge, CB3 0HA, UK
             \and
	   European Southern Observatory, Casilla 19001, Santiago, Chile
             \and
	   Institut d'Astrophysique et de G\'{e}ophysique, Universit\'{e} de Li\`{e}ge, Belgium  
	     \and
	   Ecole Normale Sup\'{e}rieure, Paris, France 
	     \and
           CEA/DSM/DAPNIA, Service d'Astrophysique, Saclay, F-91191 Gif sur Yvette, France
	     \and
	   INAF IASF Milano, via Bassini 15, 20133 Milano, Italy 
	     \and
           Astrophysics group, Blackett Laboratory, Imperial College, London, UK
          }
\offprints{P.~Gandhi \email{pg@ast.cam.ac.uk}}
\date{}

\label{firstpage}

\abstract{
We analyze X-ray sources detected over 4.2 pseudo-contiguous sq. deg. in the 0.5--2 keV and 2--10 keV bands down to fluxes of $2\times 10^{-15}$ and $8\times 10^{-15}$ erg s$^{-1}$ cm$^{-2}$ respectively, as part of the \xmm\ Large Scale Structure Survey. The \lognlogs\ in both bands shows a steep slope at bright fluxes, but agrees well with other determinations below $\sim2 \times 10^{-14}$ erg s$^{-1}$ cm$^{-2}$. The detected sources resolve close to 30 per cent of the X-ray background in the 2--10 keV band. We study the two-point angular clustering of point sources using nearest neighbours and correlation function statistics and find a weak, positive signal for $\sim 1130$ sources in the 0.5--2 keV band, but no correlation for $\sim 400$ sources in the 2--10 keV band below scales of 100~arcsec. A sub-sample of $\sim 200$ faint sources with hard X-ray count ratios, that is likely to be dominated by obscured AGN, does show a positive signal with the data allowing for a large angular correlation length, but only at the $\sim 2$ (3) $\sigma$ level, based on re-sampling (Poisson) statistics. We discuss possible implications and emphasize the importance of wider, complete surveys in order to fully understand the large scale structure of the X-ray sky.
}

\titlerunning{XMM-LSS: point source properties and angular correlations}
\authorrunning{P. Gandhi  et al.}
\maketitle

\section{Introduction}

With the latest generation of X-ray satellites, \xmm\ and \c, it has become possible to easily identify active galactic nuclei (AGN) and galaxy clusters, and map out their distribution to high redshifts \citep{brandthasinger05, rosati02_araa}. We are finally in a position to answer questions such as: In what environments do AGN preferentially form? Are AGN formation and fuelling influenced by large-scale structure, or are their properties decided by factors local to the AGN and its hosting bulge alone? Is there a dependence of AGN obscuring column density on their larger-scale environment? 

Several previous AGN spatial and angular clustering measurements have been carried out in X-rays and provide a mixed picture. The \einstein\ and \rosat\ missions sampled bright and typically unobscured AGN populations, resulting in detections of moderate clustering signatures \citep[e.g., ][ and references therein]{vikhlininforman95, carrera98, fabianbarcons92}. In the optical, \citet{wake04} have found that Seyferts in the Sloan Digital Sky Survey at $z<0.2$ selected on the basis of their \oiii\ or \nii\ emission line strengths are unbiased tracers of mass, with neither their auto-correlation properties, nor cross-correlation with galaxies showing significant excess above the field. How these results extend to high redshift and connect with AGN selected at other wavelengths is a subject of intense study. 

More recent measurements between 2 and 10 keV are capable of probing through increasing columns of absorbing material associated with the tori of obscured AGN. Since obscured AGN outnumber their unobscured counterparts by a factor of anywhere between 3 and 10 \citep{maiolinorieke95,matt00} and since X-ray surveys select AGN much more efficiently than at other wavelengths \citep{brandt_chile}, hard-band studies (above $\sim$2~keV) are essential to draw conclusions from a representative AGN census. Such work has been carried out by \citet{yang03} with \c\ and \citet{basilakos04} with \xmm\ over areas covering 0.4~\sqd\ and 2~\sqd\ respectively, and both find a significant auto-correlation signal, possibly associated with the distribution of obscured AGN. 

In this paper, we describe initial results on the properties and distribution of X-ray-detected AGN in a large survey: the \xmm\ Large Scale Structure survey\footnote{http://vela.astro.ulg.ac.be/themes/spatial/xmm/LSS/} \citep[hereafter XMM-LSS; ][]{pierre_lss}. This is a contiguous, wide-area (currently $\sim$6~deg$^2$) survey with the primary goals of studying the physical properties of cluster/group populations; the impact of environment on star, AGN and galaxy formation; and, reciprocally, the effect of star formation activity on cluster properties. This is currently the widest, medium-deep survey of a contiguous patch of the X-ray sky with spatial resolution better than 10~arcsec above 2~keV. Full characterization of the nature of the detected X-ray population will be possible with extensive multi-wavelength follow-up currently under way. In addition, the XMM-LSS area and sub-regions have already been (or will be) observed as part of numerous large and \lq legacy\rq\ surveys at other wavelengths, including the radio \citep[VLA; ][]{cohen03}, optical \citep[CFHT, VLT; ][]{lefevre04}, near-infrared (UKIDSS\footnote{http://www.ukidss.org/}) and mid- to far-infrared wavelengths \citep[Spitzer; ][]{lonsdale03}. Initial follow-up of cluster candidates has proven highly successful. Results include the confirmation of several high-redshift clusters at $z>0.6$ \citep{andreon05, valtchanov04}, extension of the lower-redshift ($0.3<z<0.6$) sample to the luminosity regime of poor groups and clusters \citep{willis05,willis05erratum} as well as compilation of the highest sky density cluster sample to date \citep{pierre06}.

While the full XMM-LSS cluster survey is expected to provide sensitive measurements and consistency checks of cosmological parameters \citep{refregier02}, more than $\sim 80$ per cent of the X-ray sources detected to the flux limits of the survey are point sources, predominantly AGN. We present the basic properties of the detected point-like sources in the XMM-LSS field, including distributions of source flux above specific sensitivity limits (\S~\ref{sec:lognlogs}) and a measurement of the resolved fraction of the X-ray background (\S~\ref{sec:xrb}). This is followed by an analysis of the projected two-point correlations on the sky of various sub-samples of point-like sources (\S~\ref{sec:acf}), including a sample with hard count ratios most likely dominated by obscured AGN. Lastly, we compare these results with previous work and discuss possible implications (\S~\ref{sec:discussion}). While some source classification and separation is discussed in \S~\ref{sec:classifications}, detailed identification and follow-up (still in progress) will be presented in future works. 

Analysis of these sources over a sub-region of guaranteed-time pointings (hereafter referred to as the \lq G pointings\rq) covering $\sim 3$ sq~degs has been described by \citet{chiappetti05}. 
The present paper is an extension of these results to cover the point-sources detected in the full area of the XMM-LSS pointings observed so far, and to study their two-point angular correlations. We include an additional 30 pointings (hereafter referred to as the \lq B pointings\rq) observed as part of guest observer time. Additionally, the source detection pipeline used for our work is different from that of \citeauthor{chiappetti05} While these last authors use well-tested, standard XMM-SAS point-source detection algorithms ({\tt eboxdetect, emldetect} etc.), the main driver of the XMM-LSS survey to detect faint groups and clusters has motivated the development of a custom-built wavelet technique with a full profile fitting algorithm in order to best distinguish between point-like and extended sources \citep{pacaud06}, large parts of which we utilize in the present work. 

Throughout this paper, the hard band refers to the X-ray 2--10 keV band; the soft band to 0.5--2 keV; and the term \lq hard-spectrum\rq\ sources refers to sources with a hardness ratio of X-ray counts (\hr; the relative excess of hard band counts over the soft band) greater than --0.2. Where required, we use the concordance cosmology \citep{wmap}, unless otherwise stated.

\section{The sample of point-like sources in the XMM-LSS}

\begin{figure*}
  \begin{center}
    \includegraphics[angle=0,height=20cm]{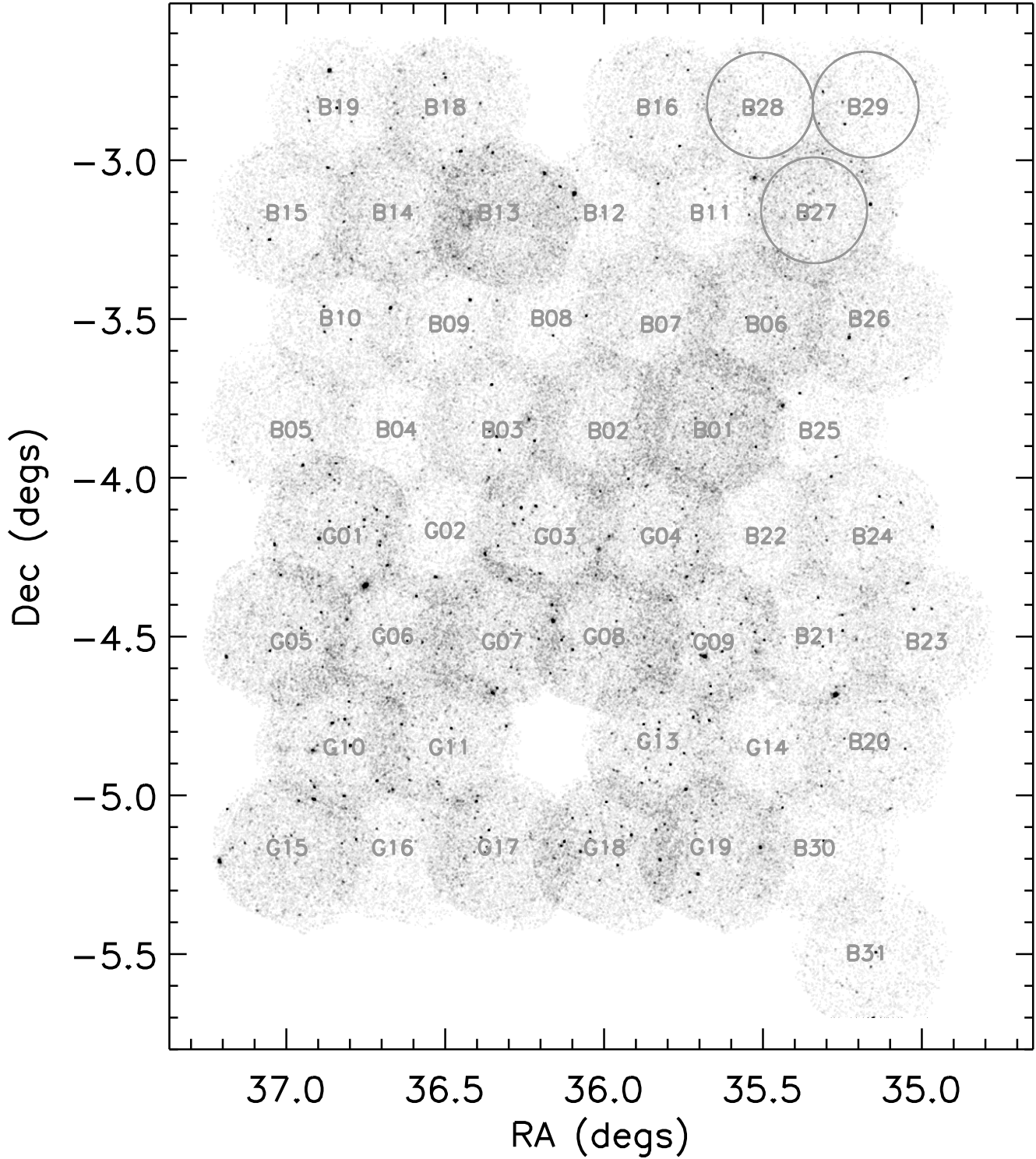}
  \caption{The layout of the 48 pointings of the XMM-LSS survey. Smoothed 0.5--10 keV photon images for all cameras have been coadded with the same scaling in counts for display (without exposure map correction). Each pointing has a full field-of-view diameter of just under 30~arcmin. The placement is such that most adjacent pointings overlap beyond $\sim$10~arcmin from the respective optical axis centres (the pointings are labelled with their field names [compare with Table~\ref{tab:fields}] at approximately these centres). In the present work, we restricted the analysis to the respective central 10$'$ regions, outlined for three pointings only (for clarity) at the top right as the 10$'$--radius large circles. The limiting sensitivity varies as a function of the exposure times, background level and off-axis angles. Many of the brighter X-ray sources are easily discerned in black. The longer-exposed guaranteed time pointings (G) are in the southern part of the survey; the rest (B pointings) being obtained during guest observer time.
\label{fig:layout} } 
  \end{center}
\end{figure*}

The XMM-LSS observations consist of 19 guaranteed-time (G) and 32 guest-observer time (B) overlapping pointings covering a total area of $\sim 6$~\sqd. The nominal exposure times are 20~ks and 10~ks for the G and B pointings, respectively. One G pointing and two B pointings\footnote{These pointings have the following \xmm\ Observation IDs respectively: 0109520401, 0037981701 and 0147111501.} with high flaring background were not analysed for this work; the remaining 48 pointings (18 G + 30 B) are listed in Table~\ref{tab:fields}, and the layout of the fields on the sky is shown in Fig.~\ref{fig:layout}. Details of the X-ray observations are described in \citet{pierre_lss}, and complete details of the detection pipeline and source classification will be presented in \citet{pacaud06}. The unique feature of the pipeline is a custom-built maximum-likelihood profile fitting algorithm (Xanim) that runs on a list of initial detections found by first using SExtractor \citep{sextractor} on wavelet-filtered, reduced images in each pointing and energy band. This increases sensitivity towards faint, extended sources while properly accounting for Point-Spread-Function (PSF) variation with all instrumental and position--dependent effects (e.g., energy, off-axis position, bad pixels and CCD gaps). Herein, we present results obtained from pipeline runs on the 0.5-2 keV (soft) and 2-10 keV (hard) band photon images, and refer the reader to \citet{pacaud06} for the full XMM-LSS pipeline algorithm.

\subsection{Source Detection and Photometry}
\label{sec:pipeline}

\begin{table}
 \begin{center}
 \begin{tabular}{ccc}
 Field -- ObsID      & Field -- ObsID & Field -- ObsID   \\
 \hline
 G01 --     0112680101 & G18 -- 0111110401 & B15 --   0037981501    \\
 G02 --     0112680201 & G19 -- 0111110501 & B16 --   0037981601    \\
 G03 --     0112680301 & B01 -- 0037980101 & B18 --   0037981801    \\
 G04 --     0109520101 & B02 -- 0037980201 & B19 --   0037981901    \\
 G05 --     0112680401 & B03 -- 0037980301 & B20 --   0037982001    \\
 G06 --     0112681301 & B04 -- 0037980401 & B21 --   0037982101    \\
 G07 --     0112681001 & B05 -- 0037980501 & B22 --   0037982201    \\
 G08 --     0112680501 & B06 -- 0037980601 & B23 --   0037982301    \\
 G09 --     0109520601 & B07 -- 0037980701 & B24 --   0037982401    \\
 G10 --     0109520201 & B08 -- 0037980801 & B25 --   0037982501    \\
 G11 --     0109520301 & B09 -- 0037980901 & B26 --   0037982601    \\
 G13 --     0109520501 & B10 -- 0037981001 & B27 --   0037982701    \\
 G14 --     0112680801 & B11 -- 0037981101 & B28 --   0147110101    \\
 G15 --     0111110101 & B12 -- 0037981201 & B29 --   0147110201    \\
 G16 --     0111110701 & B13 -- 0037981301 & B30 --   0147111301    \\
 G17 --     0111110301 & B14 -- 0037981401 & B31 --   0147111401    \\
 \hline
 \end{tabular}
 \caption{The list of \xmm\ pointings (field label and Observation ID) considered for the present analysis. 
The nominal durations of the G (guaranteed time) and B (guest observer) pointing exposures are 20 and 10~ks respectively. Additional details at http://cosmos.iasf-milano.inaf.it/$\sim$lssadmin/Website/LSS/Anc/\label{tab:fields}} 
 \end{center}
\end{table}

For each source, the maximum-likelihood normalization of the PSF profile over the local background determines the photometry in counts, after appropriately masking out any neighbouring sources by the use of segmentation maps and accounting for chip gaps. The fit is carried out over an aperture large enough to encompass the bulk of the counts (typically a 70~arcsec box for point-sources, and a larger aperture for extended ones, depending on source extent). Conversion from count-rate ($CR$) to flux ($F$) is computed from a combination of the inverse conversion factors ($CF$) for each of the cameras, scaled by the exposure times as in \citet{baldi02}. 
The conversion factors for individual cameras were calculated using {\sc xspec} \citep{xspec} and the latest available EPIC response matrices. The thin filter response was considered, as for the actual observations. The model used was an intrinsic power-law $\Gamma = 1.7$ affected by Galactic absorption of $N_{\rm H}=2.6\times 10^{20}$ cm$^{-2}$, appropriate to the XMM-LSS sight-line. The individual $CF$s are listed in Table~\ref{tab:cfs}.

\begin{table}
\begin{center}
 \begin{tabular}{lcc}
  EPIC Camera     &      0.5--2 keV     &      2--10 keV \\
  \hline
        MOS       &      $4.990\times 10^{-12}$      &      $2.296\times 10^{-11}$ \\
         pn       &      $1.460\times 10^{-12}$      &      $7.912\times 10^{-12}$ \\
  \hline
 \end{tabular}
 \caption{The count-rate--to--flux conversion factors for the individual EPIC cameras and energy bands, stated in units of erg s$^{-1}$ cm$^{-2}$ for a rate of 1~ct~s$^{-1}$. A photon-index power-law of $\Gamma=1.7$ affected by Galactic absorption of $2.6\times 10^{20}$ cm$^{-2}$ was assumed. 
Both MOS cameras were assumed to be identical.\label{tab:cfs}}
\end{center}
\end{table}

\subsection{Source selection and classification}
\label{sec:classifications}

Intensive follow-up of \c\ and \xmm\ surveys has shown that the vast majority of X-ray detections at high galactic latitudes are associated with AGN, in deep as well as in shallow surveys \citep{brandt_chile, szokoly04, nandra04}. The optical identifications of the associated counterparts are varied, and include Seyferts as well as late and early-type galaxies. Despite the fact that these do not always show obvious optical signs of AGN activity, nevertheless the presence of a powerful accretion source is indisputable, based on the inferred power of the X-ray source, and, in many cases, the detection of weak, high-ionization emission lines. 
In the XMM-LSS survey as well, the bulk (more than $\sim 80$ per cent) of X-ray detections will be due to AGN (compare with, e.g., \citealt{c02}). Stars and bremsstrahlung emission from galaxy clusters will constitute the remaining sample. 

Obvious stars were removed by cross-correlating our X-ray detections with the USNO survey \citep{usno} and identifying all optical point-like sources with a $B$-band magnitude brighter than 13. 
This cross-correlation resulted in 17 sources being removed from the 0.5--2 keV sample, and these are not included in any of the following analysis. No stars coincided with a hard-band X-ray source above the nominal significance threshold (see below). Though there will undoubtedly remain some contamination due to fainter non-AGN point-like sources, the \rosat\ deep surveys have found that stars represent only a minority of all X-ray detections \citep{schmidt98} and typically display soft X-ray spectral slopes. Thus, our results should not be affected significantly by these contaminants, especially in the hard-band. 

In the process of identifying clusters and groups (hereafter, jointly referred to as clusters), extensive simulations are being carried out to determine the best X-ray classification parameters, based on already published clusters \citep[][ amongst others]{valtchanov04, willis05,willis05erratum}. For the purpose of the present work, we take as \lq extended\rq, only those soft-band (0.5--2 keV) sources that have already been followed-up at other wavelengths and confirmed to be clusters, or found to have a high probability of being associated with such systems, based on i) optical-spectroscopy; ii)  X-ray source profile extent; and, iii) the presence of a red sequence of galaxies (see \citealt{pacaud06} for full details). At the time of writing, this consists of 28 confirmed clusters, and 28 clusters with provisional spectroscopic redshifts over the full 5.7~\sqd\ of Fig.~\ref{fig:layout}. The rest of the sources in the soft band are classified as \lq point-like\rq. In the 2--10 keV (hard) band, we treat all sources as being point-like.

For the analysis presented in this paper, only those sources that lay within 10~arcmin of the optical axis centres of each pointing were retained. This was done in order to minimize biases due to PSF distortion at large off-axis angles and possible confusion due to the same source being detected on adjacent pointings. This results in pseudo-contiguous coverage of the field, with holes in between neighbouring pointings (that are fully accounted for in our analysis; see below). 

Finally, the significance of source detection was estimated by the value of signal:noise (S/N). The \lq signal\rq\ used is that corresponding to 68 per cent of the full background-subtracted counts for each source, and the \lq noise\rq\ assumes Poisson errors on the total counts (source+background) within a fixed aperture of radius of 17 arcsec. Strictly speaking, this is consistent only for on-axis point-sources\footnote{http://xmm.vilspa.esa.es/external/xmm\_user\_support/document-ation/uhb/node17.html}, though the effect of the off-axis PSF degradation is mitigated somewhat by our restriction to the central 10~arcmin regions of each pointing. To compute the Poisson error, we specifically use Eq. 7 of \citet{gehrels86} for the 1$\sigma$ upper-limit on the noise. Most results in the following sections are presented for the sample with S/N$>$3, but we also consider an effectively-fainter sample with S/N$>$2 in \S~\ref{sec:hardsources}, and briefly mention results relevant to sub-samples with higher S/N thresholds of 4 and 5.

\begin{table}
 \begin{center}
 \begin{tabular}{llcc}
   Selection Criterion   &  Classification          & 0.5--2 keV   &  2--10 keV\\
   \hline
   S/N$>$3 (B+G)         &     Point sources        &       1134   &     413   \\
   ~~~~~~~~$''$          &  Extended sources        &        36    &     0$^a$ \\
   ~~~~~~~~$''$          &             Stars        &        17$^b$&     0     \\
   S/N$>$2 (B+G)         &     Point sources        &        --    &     912   \\
   S/N$>$2 (G)           &     Point sources        &        --    &     473   \\
   ~~~~~~~~$''$          &   1$\ge$\hr$>$--0.2      &        --    &     209   \\
    ~~~~~~~~$''$         &   1$>$\hr$>$--0.2        &        --    &     140   \\
   \hline
 \end{tabular}
 \caption{Sizes of various sub-samples of X-ray sources analyzed in this paper. For the S/N threshold of 2, only hard-band sources are analyzed in \S~\ref{sec:hardsources}. The parentheses in the first column specify whether the sample was selected over the whole area (B+G pointings), or only over the deeper G pointings. Notes: $^a$All sources were considered as point-like in the 2--10 keV band. If we relax this condition, only 8 sources are affected.
$^b$Only obvious stars with $B<13$ were identified (see text), and these are counted separately from \lq point sources\rq.\label{tab:sourcenumbers}}
 \end{center}
\end{table}

\subsection{Results: Sky coverage and log$N$--log{S}}
\label{sec:lognlogs}

\begin{figure*}
  \begin{center}
    \includegraphics[angle=90,width=8.6cm]{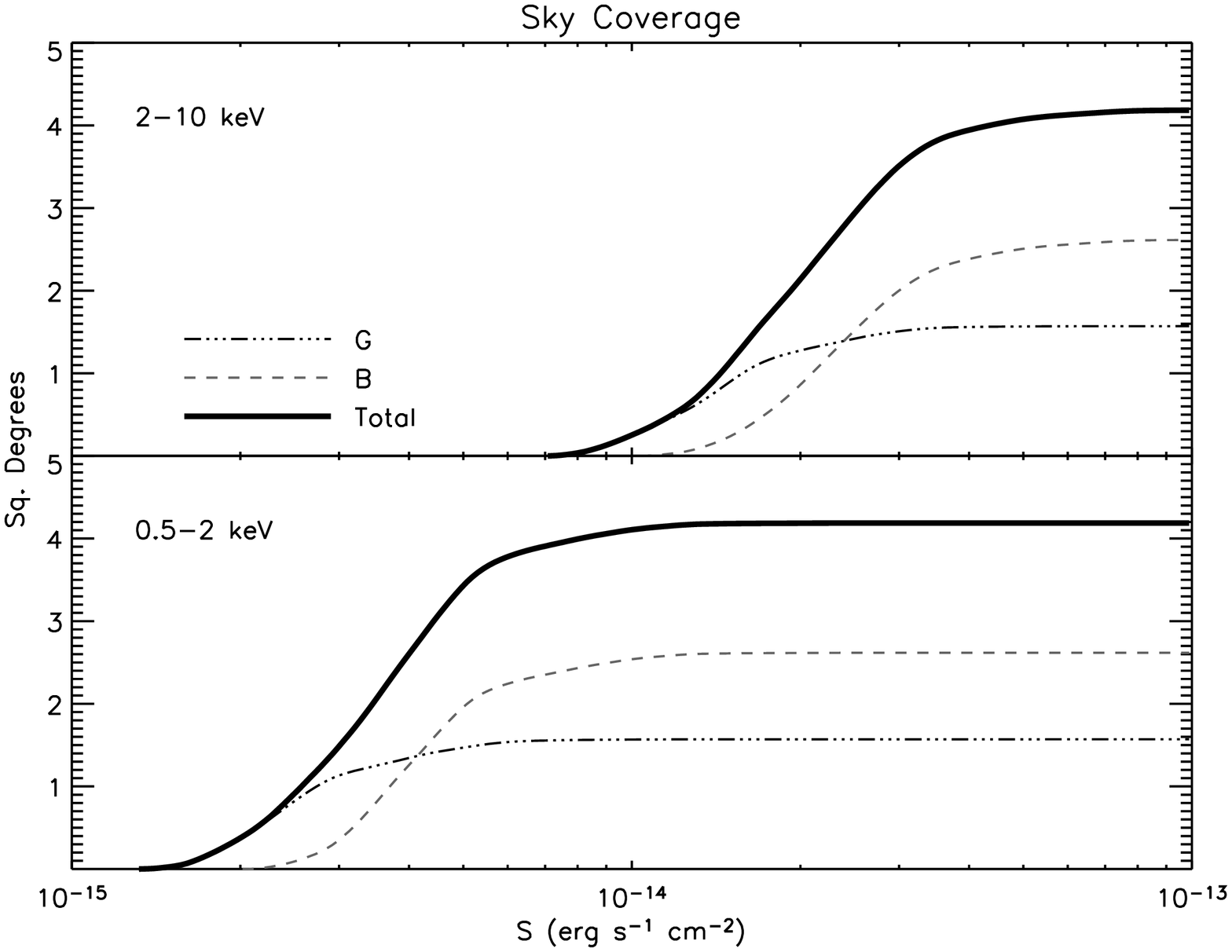}
    \includegraphics[angle=90,width=8.6cm]{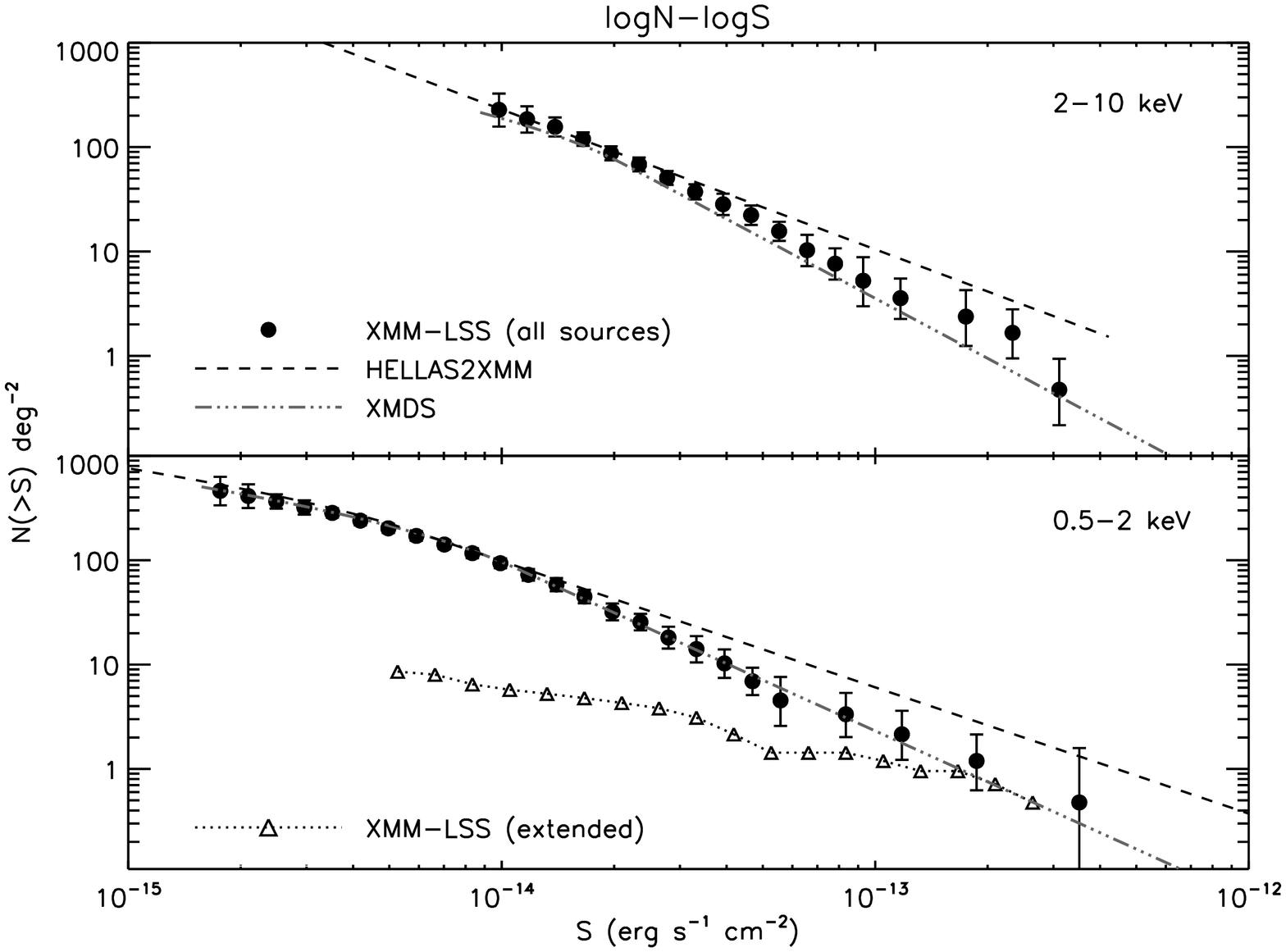}
  \caption{The sky coverage ({\em left}) and \lognlogs\ ({\em right}) of the XMM-LSS sample within the central 10~arcmin-radius pointing regions, for the 2--10 keV ({\em top}) and 0.5--2 keV ({\em bottom}) bands, for a threshold S/N$>$3 in both bands. The sky coverage is shown separately for the guaranteed time (deeper; dots-dashed; marked \lq G\rq) and guest observer time (shallower; dashed; marked \lq B\rq) pointings. The \lognlogs\ is shown for all sources (clusters have a minor contribution, except at bright fluxes in the soft band; their contribution is shown as the triangles; marked \lq extended\rq). The errors on the \lognlogs\ denote 1$\sigma$ Poisson uncertainties on the independent differential count bins, subsequently scaled to the integral counts. The results of the HELLAS2XMM survey \citep{baldi02} and \xmm\ Medium Deep Survey \citep[XMDS; ][]{chiappetti05} are also overplotted for comparison. 
\label{fig:coverage_lognlogs}}
  \end{center}
\end{figure*}

Sensitivity maps for each pointing were constructed by using the background measured by the source detection pipeline. The minimum number of counts necessary at any spatial pixel on a pointing (above the local background) in order for a source to have a S/N matching the adopted threshold is computed, and converted to a limiting-flux using the exposure maps and conversion-factors. The sky coverage is computed by summing the area covered by these sensitivity maps as a function of limiting-flux, and is shown for the hard and soft bands in Fig.~\ref{fig:coverage_lognlogs}, for the nominal S/N threshold of 3. The longer-exposed guaranteed time (G) pointings have a deeper sensitivity, but smaller overall coverage than the guest observer (B) pointings. The maximum coverage at the brightest fluxes is close to 4.2 \sqd; overlap between adjacent pointings in this selected area of the central 10~arcmin regions is negligible (only 0.3 per cent of the total).

Above a S/N threshold of 3, we find a total of 1134 and 413 point sources in the soft and hard bands respectively (Table~\ref{tab:sourcenumbers} lists the numbers of sources in various bands and with different selection criteria as discussed in the text). The \lognlogs\ in each band (also shown in Fig.~\ref{fig:coverage_lognlogs}) was computed by identifying all sources with a detected flux above any given value $S$ and summing the inverse areas over which these sources could have been detected, as measured from the sky coverage. The flux distributions of the detected sources peak at $\sim 5\times 10^{-15}$ and $\sim 1.5\times 10^{-14}$ \ergpspsqcm\ in the soft and hard bands respectively, and at these flux levels, the \lognlogs\ is in excellent agreement with other published surveys (e.g., we show the fit of \citealt{baldi02} in Fig.~\ref{fig:coverage_lognlogs}). The faintest fluxes detected (typically near the optical axes) are $\sim 2$ and $8\times 10^{-15}$ \ergpspsqcm\ in the two bands, respectively. At the lowest fluxes in the soft band, our \lognlogs\ begins to drop off due to incompleteness, while at bright fluxes in both bands, our \lognlogs\ is lower than (but consistent with the 1$\sigma$ lower limit of) other surveys. This difference was also observed by \citet{chiappetti05} over the area of the G pointings, based on a completely different source detection procedure (their \lognlogs\ is also shown in Fig.~\ref{fig:coverage_lognlogs}). We find excellent agreement with \citet[][ especially in the soft band]{chiappetti05} and refer to their paper for power-law fits and discussion of the \lognlogs. 
The agreement gives us confidence that the noted deficit is intrinsic and not due to pipeline systematics.

\subsection{Results: Resolved fraction of the X-ray background}
\label{sec:xrb}

Only with the new generation of X-ray satellites has the hard cosmic X-ray background radiation been resolved substantially into discrete sources (a combination of obscured and unobscured AGN), supporting the basic tenet of AGN Unification \citep{settiwoltjer89, mushotzky00}. While the soft X-ray background below 2 keV can be accounted for almost completely as a combination of emission from Galactic emission, AGN and clusters \citep[e.g., ][]{fabianbarcons92}, the exact fraction resolved out in the hard band remains a contentious issue. There is even intriguing evidence that a new population of obscured AGN remains undiscovered in even the deepest surveys \citep{delucamolendi04, worsley04}.

\begin{figure}
  \begin{center}
    \includegraphics[angle=90,width=8.5cm]{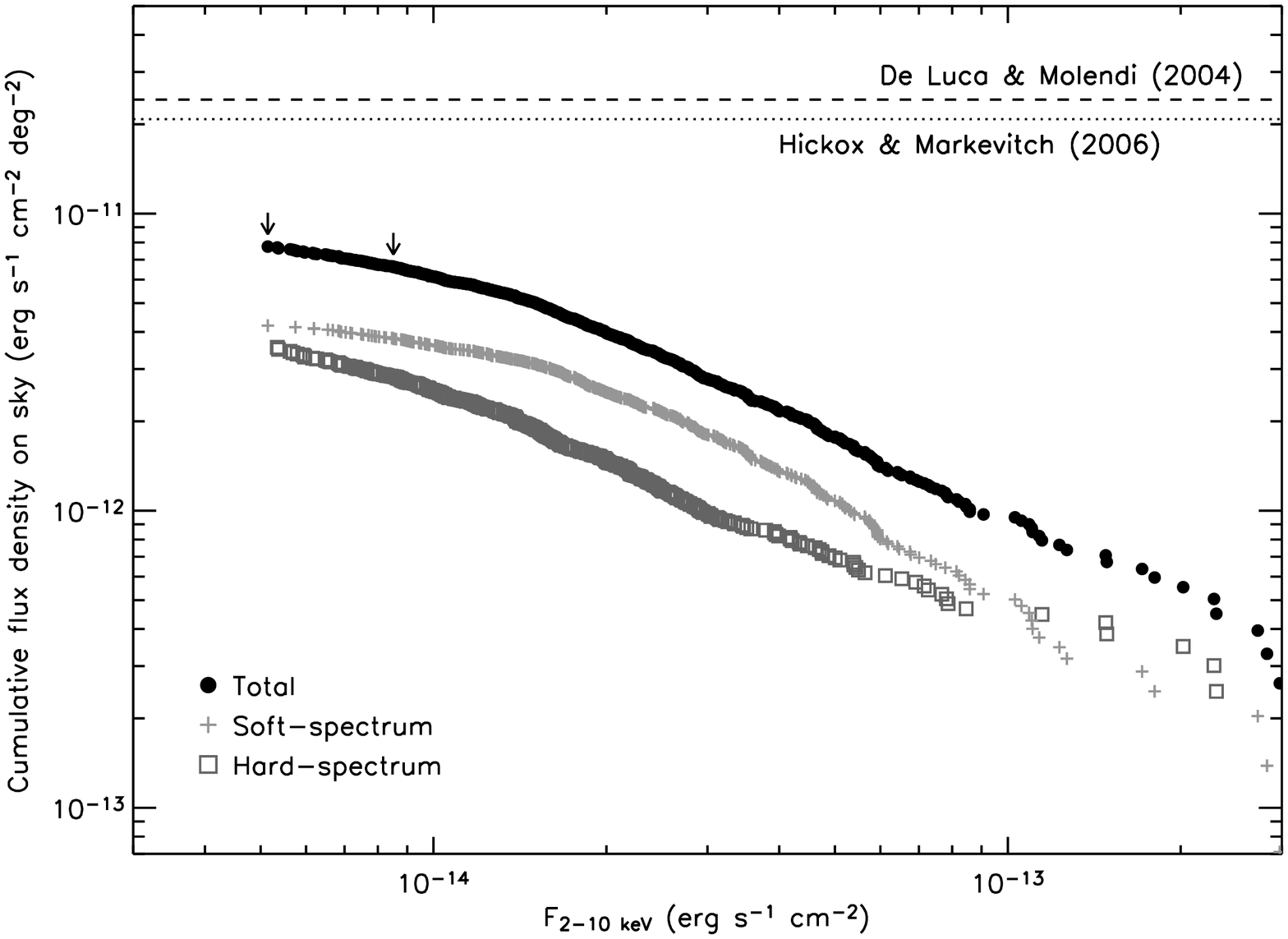}
  \caption{Cumulative X-ray background intensity in the 2--10 keV band for the detected point sources in the full field of the XMM-LSS survey. Arrows denote the faintest flux limits of sub-samples compiled with S/N thresholds of 3 ({\em right arrow}) and 2 ({\em left arrow}), respectively. The dashed line is the total X-ray background measurement reported by \citet{delucamolendi04}; the dotted line is the measurement of \citet{hickoxmarkevitch06} converted to 2--10 keV assuming a power-law with photon index $\Gamma=1.4$. The resolved source contribution is sub-divided into soft-spectrum and hard-spectrum sources (discussed in \S~\ref{sec:hardsources}).
\label{fig:xrbintensity}}
  \end{center}
\end{figure}

We compute the resolved intensity of detected sources by summing over the flux of each source dividing by the inverse area over which the source would have been detected, similarly to the computation of the \lognlogs. The result is shown in Fig.~\ref{fig:xrbintensity} and tabulated in Table~\ref{tab:xrbintensity}. At the nominal flux limit of the XMM-LSS survey (with S/N$>$3), we resolve close to 30 per cent of some latest measurements of the 2--10 X-ray background \citep{delucamolendi04, hickoxmarkevitch06}. This matches well with the level resolved in other surveys over our flux regime \citep[e.g., ][]{manners03}. Note that hard-spectrum sources (discussed further in \S~\ref{sec:hardsources}) contribute a larger fraction at fainter fluxes due to their increasing dominance, relative to those with softer X-ray spectra. 

\begin{table*}
 \begin{center}
 \begin{tabular}{cccc|ccc}
S/N threshold & Faintest flux & Median Flux & XRB sky intensity & \multicolumn{3}{c}{Resolved fraction (per cent)}\\
              &  erg s$^{-1}$ cm$^{-2}$ &  erg s$^{-1}$ cm$^{-2}$ &  erg s$^{-1}$ cm$^{-2}$ deg$^{-2}$ &  All sources & Hard-spectrum sources & Soft-spectrum sources \\
 \hline
S/N $>$ 3 & $8.5\times 10^{-15}$ & $2.6\times 10^{-14}$ & $6.6\times 10^{-12}$ & 28 & 12 & 16\\ 
S/N $>$ 2 & $5.1\times 10^{-15}$ & $1.8\times 10^{-14}$ & $7.7\times 10^{-12}$ & 33 & 16 & 17\\ 
 \hline
 \end{tabular}
 \caption{Resolved fraction of the X-ray background over the B+G pointings for sub-samples with different S/N thresholds. The resolved fraction is stated as a percentage of the total intensity of $2.4\times 10^{-11}$ erg s$^{-1}$ cm$^{-2}$ deg$^{-2}$ found by \citet{delucamolendi04}. The resolved fraction is further sub-divided into the contribution of hard-spectrum and soft-spectrum sources (see \S~\ref{sec:hardsources}).\label{tab:xrbintensity}} 
 \end{center}
\end{table*}

\section{Angular Correlation Statistics}
\label{sec:acf}

Full clustering analysis requires {\em spectroscopic} redshifts, since even optimistic photometric redshift errors of $\Delta z\sim 0.1$ translate into typical physical separations of hundreds of Mpc, washing out any intrinsic clustering signal. Until the on-going multi-wavelength follow-up of X-ray sources produces a useful number of accurate redshifts, we restrict our study to the areal distribution of sources only. 

Angular clustering is related to the excess probability of finding source pairs at any given angular separation $\theta$, relative to a sample distributed with uniform probability. A variety of methods have been used for studying clustering properties of astronomical surveys, including power spectrum \citep{psa}, counts-in-cells \citep[e.g.,][]{carrera98} and fractal analyses \citep{clustering_fractals}, each of which can be implemented either in real, or in projected space. Here, we describe results obtained from nearest-neighbours and correlation function statistics for the point-like sources in the XMM-LSS.

\subsection{Generation of random (uncorrelated) catalogues}
\label{sec:randomcatalogues}

The most straightforward way to account for instrument and pipeline selection effects in the survey (e.g., off-axis vignetting, holes between adjacent pointings, chip gaps) is to simulate an ensemble of catalogues over randomly-chosen sky positions that also correctly accounts for these effects. Source fluxes were randomized to have an overall distribution similar to the data. A random flux ($S_r$) can be sampled from the data \lognlogs\ by choosing a random number ($p$) uniformly distributed between 0 and 1, and searching for the flux $S_r$ that satisfies the following transformation: 

\begin{equation}
N(>S_r) = N(>S_{lim}) \times [1-p]
\end{equation}
where $N(>S_r)$ is the value of the \lognlogs\ at $S_r$ and  $S_{lim}$ is the limiting flux above which the \lognlogs\ is defined \citep[see also ][]{mullis04}. Sky positions for these random catalogues were assigned with equal probability within the central 10$'$ regions of the pointings. If the limiting-flux at the assigned coordinates was larger (worse) than the randomly-chosen flux for the source, the source was discarded, and another one randomized in its place. We verified that the \lognlogs\ distributions of the random catalogues were similar to those of the parent catalogues used to generate them in each band and with each selection criterion. 

This procedure assumes that a source with a randomly generated flux $S_r$ would have been detected with exactly that flux at the assigned sky position. In reality, any detection procedure will introduce shot noise, which results in some faint sources below the nominal threshold still being detected due to Poisson uncertainties \lq boosting\rq\ their counts. On the other hand, some bright sources may have their counts \lq depressed\rq\ and thus lie below the detection threshold. We simulated this by converting the flux $S_r$ to counts (with the use of local exposure maps and conversion-factors) and drawing a random number from the Poisson distribution of the total counts (source+local background). 
The source is then retained in the final random catalogue with a probability that the Poisson distribution exceeds the total limiting counts required by the S/N threshold. 
This simulation of shot noise was used for all random catalogues in each of the bands and sub-samples tested.

The above manual introduction of shot noise was necessary because full simulations of the entire dataset proved to be unrealistic. Still, as a cross-check, we did carry out full simulations of three pointing exposures by randomly generating sources with appropriate off-axis and energy-dependent PSFs for each camera, along with simulated backgrounds in accordance with \citet{readponman03}, all scaled for the respective pointing exposure times. Our full detection pipeline was then run over these. 
 We verified that the statistical distribution of sources over these simulated pointings compares well with the random catalogues above.

With an \xmm\ Mirror Module PSF characteristic size of $\sim 6$~arcsec, there is a small, but finite probability that blending of two close, neighbouring sources could result in them being classified as a single, extended source. The XMM-LSS pipeline has been designed to optimally detect and probe the evolution of extended sources, and we thus make use of its strength of separating point-like sources from extended clusters. We studied the effect of blending through several sets of simulations and confirmed that our pipeline is able to resolve close-by pairs successfully, though the efficiency of doing so depends on source flux as well as on pair separation. While no sources will be resolved at separations of less than $\sim 6$~arcsec, the pipeline resolves effectively all source pairs at separations of $\ge 30$~arcsec. 
Based on the simulations and observed source counts, the efficiency of pair resolution at a separation of $\sim$20 (10)~arcsec was determined to be $\sim$70 (30) per cent. Further details will be presented in \citet{pacaud06}, but for our purposes, we simply remove a corresponding fraction of such close pairs from the random catalogues in order to simulate this blending.

A hundred simulated catalogues are generated for each band and sub-sample studied, each with the same number of sources as the parent data catalogue. The average number of data and random pairs in the ensemble are then counted for the computation of correlation statistics.

\subsection{Results: Angular Nearest Neighbours Statistic}

\begin{figure}
  \begin{center}
    \includegraphics[angle=90,width=8.5cm]{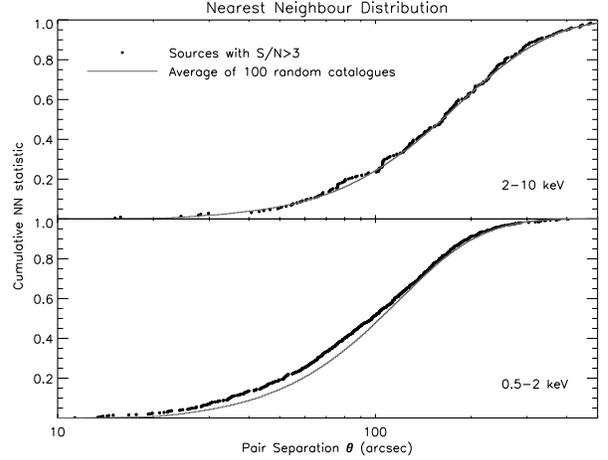}
  \caption{Cumulative nearest-neighbour distribution function (statistic) for the soft ({\em bottom}) and hard ({\em top}) bands for the point-sources with S/N$>$3 (filled circles) compared to the average statistic of 100 random catalogues. The plot shows the fraction of sources with an angular separation less than any given $\theta$ to their respective nearest neighbours. The soft band distribution has an excess of pairs compared to random, as opposed to the hard band sample.\label{fig:nn}} 
  \end{center}
\end{figure}

First, we show the distribution of projected separations of point-like sources from their first nearest neighbours (NN). This is shown as the cumulative (normalized) distribution in Fig.~\ref{fig:nn}, compared to the average distribution of nearest neighbours for 100 random catalogues, in both the soft and the hard bands. An excess of nearest neighbours is observed in the soft band ({\em bottom plot}) below $\sim$100~arcsec. A Kolmogorov-Smirnov (K-S) test returns a small probability of $\approx 10^{-3}$ for the null hypothesis that the two distributions are identical, implying possible clustering in this band. In the hard band, no such excess is observed, and the K-S probability is consistent with the data and random distributions being drawn from the same population.

\subsection{Results: Angular Correlation Function}

Optimal estimators are widely used to quantify the overall excess of data-data pairs over random-random ones at different scales: we chose to use the Hamilton estimator \citep{hamilton93} for the angular correlation function (ACF), but found very similar results using others, e.g., that of \citet{efstathiou91}. Excess clustering compared to a uniform distribution is parametrized in terms of $\omega(\theta)$, defined as 
\begin{equation}
\omega(\theta)=f\frac{DD(\theta)RR(\theta)}{DR(\theta)DR(\theta)}-1
\end{equation}
where $DD$, $RR$ and $DR$ are the number of data-data, random-random and data-random pairs at separation $\theta$, all subjected to the survey selection effects. The normalizing factor $f$ is $4 N_D N_R/(N_D-1)(N_R-1)$, where $N_D$ and $N_R$ are the number of sources in the data and random catalogues, respectively. Source pairs were binned in equal logarithmic intervals of $\theta$: the bin sizes being chosen to include at least $\sim$20 pairs in each bin in order to minimize the effect of small-number statistics. This restriction also defined the minimum pair-separation bin over which $\omega$ is plotted and fitted: this typically lies between 20--50~arcsec. In any case, the overall results presented below are not sensitive to reasonable binning choices.

\begin{figure}
    \includegraphics[angle=90,width=8.5cm]{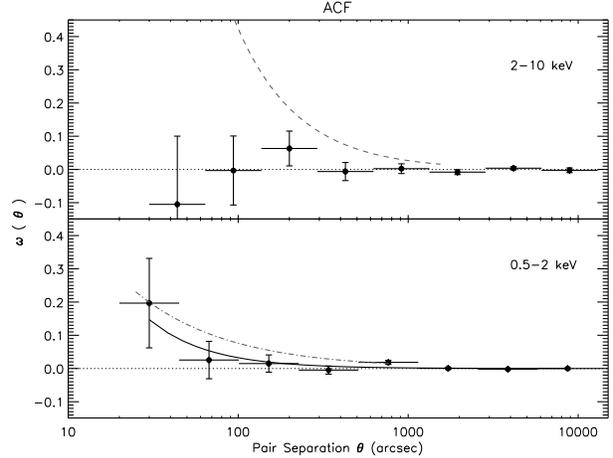}
  \caption{The ACF, as defined by \citet{hamilton93} and measured for the XMM-LSS survey in the soft ({\em bottom}) and hard ({\em top}) bands for the samples with S/N$>$3. 
The solid curve is the best-fit power-law model (shown for the soft band only), while the dotted line marks $\omega=0$. The y-axes are plotted on a linear-scale to aid visualization of the significance of any correlation and the axes ranges are kept the same in both bands for comparison. Previous power-law ACFs of \citet[][ for the hard band]{basilakos04} and of \citet[][ for the soft band]{vikhlininforman95} are shown as the dashed and dot-dashed lines respectively. \label{fig:acf}} 
\end{figure}

We show the ACF results for point-like sources in Fig.~\ref{fig:acf}. 
The plotted error-bars are Poisson 1-$\sigma$ uncertainties, calculated for each bin as $(1+\omega)/\sqrt{DD}$. In the soft band, there is an overall positive auto-correlation signal at most scales smaller than $\sim$1000 arcsec. The amplitude of this correlation is small: if we characterize $\omega$ as a power-law of the form
\begin{equation}
\omega(\theta)=(\theta_0/\theta)^{\gamma-1}
\label{eqn:acf_pl}
\end{equation}
we find $\theta_0=6.3\pm 3$ arcsec, with a slope of $\gamma=2.2\pm0.2$, with the quoted errors being appropriate to 68 per cent confidence intervals for one interesting parameter. By simply counting the excess number of data-data pairs compared to random-random ones with all separations less than, say, 200~arcsec, we find an excess at the 2.3$\sigma$ level. This result was measured for the sample of 1134 point-like sources alone. The auto-correlation signal is almost identical for the combined catalogue of 1170 soft-band detections, including the 36 extended sources, suggesting that there is no obvious and strong cross-correlation signal between the extended and point-like samples; however, this will be studied in detail once a larger, complete sample of extended sources is compiled.

In terms of numbers of pairs, we detect 58 (251) independent data pairs with separations of $< 100\ (200)$~arcsec, while the random catalogues contain 58 (230) pairs over the same scales, on average. This result is consistent with the null hypothesis implied by the nearest neighbour distribution. Though statistics are small, we find similar null results for sub-samples of hard-band sources selected with S/N$>$4 (209 sources) or with S/N $>$5 (123 sources).

We note that bias (underestimation of $\omega$) related to the integral constraint (the fact that a finite sky area with an unknown source density is used to estimate the correlation signal) is negligible for our field. Assuming the power-law form of Eq.~\ref{eqn:acf_pl} for the intrinsic correlation function, the bias can be estimated by numerical integration of 
$\theta_0^{\gamma-1}/\Omega^2\int{\int{\theta^{1-\gamma} d\Omega_1 d\Omega_2}}$
over the whole area ($\Omega$) of the survey. For a range of relevant power-law parameters, we find this bias to be $\sim 0.01$, a quantity sufficiently small to ignore during fitting.

\subsection{Results: Sources with hard X-ray spectral indices}
\label{sec:hardsources}

The above result on the non-detection of clustering in the 2--10 keV band seems to be at odds with the findings of \citet{yang03} and \citet{basilakos04}. In both these works, excess clustering was detected in the hard band, and associated with the clustering of obscured AGN, which should constitute a larger fraction of the hard-band detections as compared to the soft band. We have made predictions for the fraction of AGN detected in the XMM-LSS survey that are expected to be obscured (see \S~\ref{sec:hardsources_discussion} below) and find that, at the flux limits probed by our sample, obscured AGN will represent approximately 30--40 per cent of the hard band detections, and about 10 per cent of the soft band sample; this fraction increases with decreasing flux levels. Obscured AGN can be efficiently (but not uniquely) selected by computing the hardness ratio (\hr) of source counts in the hard ($H$) and the soft ($S$) bands. If we define
\begin{equation}
HR=\frac{H-S}{H+S},
\end{equation}
then a large fraction of sources with \hr$>$--0.2 (hereafter, \lq hard-spectrum sources\rq) are likely to be obscured AGN. This limit corresponds to an obscuring column density of approximately $10^{22}$~cm$^{-2}$ due to cold gas local to a source with an intrinsic power-law photon index of 1.7 at a redshift $z=0.7$, and has been often used in the literature to separate intrinsically obscured AGN from unobscured ones \citep[e.g., ][]{g04, padovani04}. 

Applying the above \hr\ criterion to the hard band sample of 413 sources results in only 133 hard-spectrum sources over the B+G pointings, which is not sufficient for a proper correlation analysis. Since obscured AGN begin to emerge at the faintest fluxes, it is possible to probe them in larger numbers by decreasing our significance threshold for source detection. We therefore searched for detections with S/N$>$2, and found a total of 912 sources in the 2--10 keV band, of which 409 have \hr$>$--0.2. But only a marginal correlation signal was detected for these. 

\begin{figure}
  \begin{center}
    \includegraphics[angle=90,width=8.5cm]{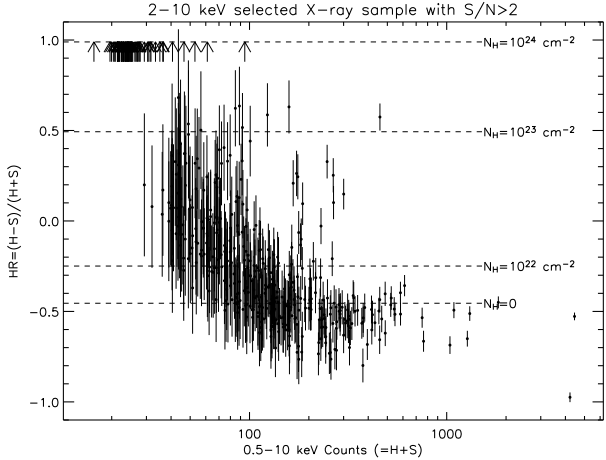}
  \caption{Hardness ratio (\hr) vs counts (Hard+Soft; i.e. $H+S$) of the G pointings sample of sources with S/N$>$2 in the 2--10 keV band with either a unique counterpart (400 black dots with error bars) or no counterpart (69 arrows) in the soft band. 
The horizontal dashed lines mark the hardness ratios of a power-law source at $z=0.7$ with photon-index $\Gamma=1.7$ observed on the pn CCD, and obscured by various columns of gas (labelled as \nh) local to the source, in addition to the constant Galactic column. 
We take all sources with \hr$>$--0.2 to be \lq hard-spectrum\rq\ sources.\label{fig:hrcts}} 
  \end{center}
\end{figure}

We note, however, that there is a large variation in the exposure-times and limiting fluxes of the individual pointings, especially between the B and G pointings, which could bias results due to varying efficiency of selecting obscured AGN across the field. In order to have a more uniform coverage, we then restricted our source selection to the deeper G-pointings only. 
The total number of 2--10 keV detections with S/N$>$2 over the G pointings is 473. Of these, 400 have unique counterparts in the 0.5--2 keV band within a threshold inter-band distance of 10~arcsec, and 140 of these are hard-spectrum sources with \hr$>$--0.2 (Fig.~\ref{fig:hrcts}). In order to include the hardest sources, no S/N selection was imposed on the soft-band sample for this cross-correlation. Additionally, 69 sources (of 473) have no counterpart in the 0.5--2 keV band, and are likely to be very highly obscured (\nh$>$10$^{23}$~cm$^{-2}$ or, possibly, Compton-thick) sources, if they are non-spurious detections (see \S~\ref{sec:hardsources_discussion}). We combine the sub-samples of the 140 detections [with 0.5--2 keV counterparts] and of 69 detections [without counterparts], giving a total of 209 hard-spectrum sources.

\begin{figure}
  \begin{center}
    \includegraphics[angle=90,width=8.5cm]{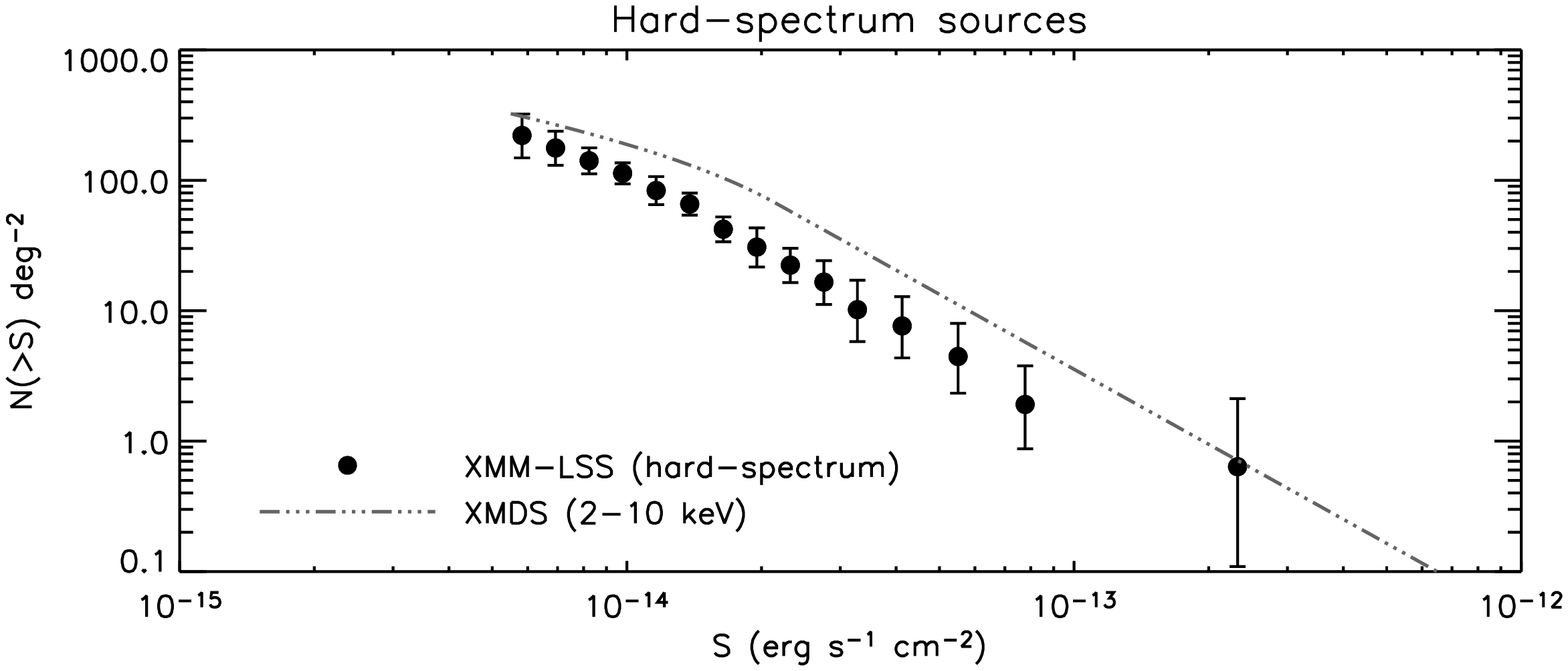}
    \includegraphics[angle=90,width=8.5cm]{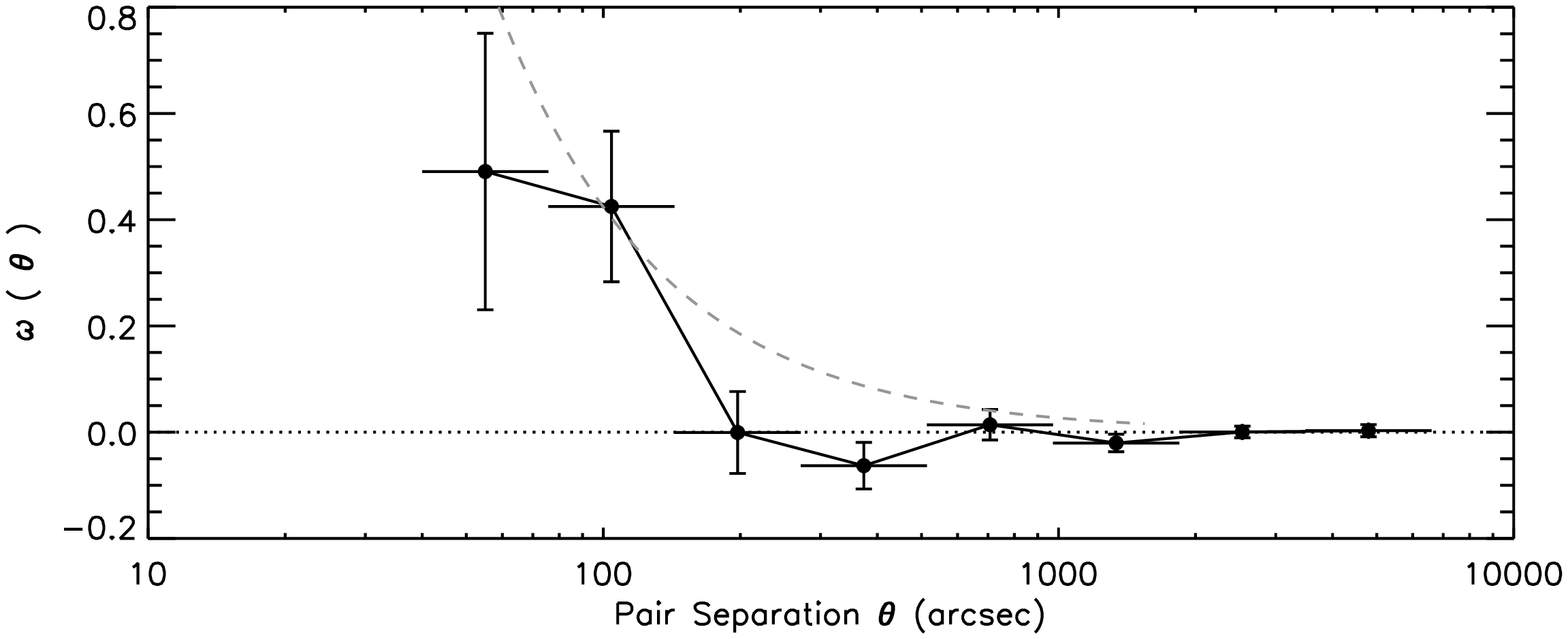}
  \caption{\lognlogs\ ({\em top}) and ACF ({\em bottom}) for the 2--10 keV sample of 209 hard-spectrum (\hr$>$--0.2) sources with S/N$>$2 in the G pointings. In the \lognlogs\ plot, the dot-dashed line shown for comparison is the best-fit to the full 2--10 keV XMDS \lognlogs, which also traces the slope of our full sample of hard-band sources 
(see Fig.~\ref{fig:coverage_lognlogs}). In the ACF plot, notice the larger range of the y-axis compared to Fig.~\ref{fig:acf}. The dashed line is the best-fit ACF of \citet{basilakos04}. \label{fig:acf_hardsources}} 
  \end{center}
\end{figure}

The nearest neighbours (NN) distribution of these hard-spectrum sources shows departure from a uniform distribution at the 99.4 per cent probability level (based on the K-S test). The \lognlogs\ and auto-correlation function (ACF) of these 209 sources is plotted in Fig.~\ref{fig:acf_hardsources}. $\omega$ is clearly positive in the first two bins, representing a 3.0$\sigma$ excess assuming Hamilton's formula (or 3.4$\sigma$ assuming that of \citeauthor{efstathiou91}). Though a power-law is not a good description of the signal, a full fit with the model of Eq.~\ref{eqn:acf_pl} results in a normalization $\theta_0=42^{+7}_{-13}$~arcsec and slope $\gamma=3.1_{-0.5}^{+1.1}$ (uncertainties are, again, those relevant for one parameter of interest based on Poisson errors). 

A similar auto-correlation analysis of the 260 sources with {\em soft} spectral count ratios (\hr$<$--0.2) shows no such excess at scales less than 100~arcsec. We also note that the correlation signal of hard-spectrum sources is not dominated by the very hardest sources alone (the arrows in Fig.~\ref{fig:hrcts}). Though the sample size is small, the NN statistic of only the 140 hard-spectrum sources {\sl with} soft-band counterparts gives a K-S null hypothesis probability of only $10^{-3}$ compared to 100 random catalogues, again suggesting departure from a uniform distribution. 

The main results of the correlation analysis on various sub-samples described above are summarized in Table~\ref{tab:acfsummary}.

\subsection{On the significance level of observed correlations}

Since the above constraints on clustering are relatively weak (a characteristic of angular clustering studies of sparse samples), 
it is pertinent to examine the significance levels quoted. 

Initial tests above computed the distribution function of nearest neighbour separations, 
and the results were found to be consistent with the strength of the angular correlation function, as estimated by simple pair-counting over relevant scales. 
The Poisson errors used in the ACF fits, however, are strictly valid only for uncorrelated data. Bootstrap re-sampling \citep{bootstrap} is widely used to assess the internal reliability of a correlated dataset. Yet, as shown by \cite{fisher94}, for sparse samples, this can over-estimate the true uncertainties by factors of $\sim 2$ (and up to 4). For uncorrelated, and weakly correlated samples, such as ours, Poisson errors approximate the true errors, despite being a lower-limit. This approximation is likely to break down for the sample of hard-spectrum sources which shows the largest deviation from a random distribution.  
In this case, we re-sampled the entire ensemble of random catalogues a large number of times ($\ge 50$) with replacement in order to compute the bootstrap errors. 
We find $\theta_0=42'' \pm 22$ and $\gamma=3.1 \pm 1.3$, implying $\sim 2\sigma$ constraints on both the normalization and the slope of the correlation function fit of the previous section (errors denote dispersion amongst the re-samples).  

Since the bins used in the ACF analysis are themselves correlated at different scales, we compute the covariance matrix Cov($\theta_i$, $\theta_j$) of $\omega$ returned by the above bootstrap method between all pairs of bins ($\theta_i$, $\theta_j$) used for the fit. The correlation matrix is then calculated; it is simply the covariance matrix scaled to the diagonal elements as follows: 

\begin{displaymath}
\hspace*{0.5cm}{\rm Corr}(\theta_i, \theta_j)={\rm Cov}(\theta_i, \theta_j)/\sqrt{{\rm Cov}(\theta_i,\ \theta_i)\ {\rm Cov}(\theta_j,\ \theta_j)}
\end{displaymath}

To estimate the strength of the off-diagonal correlations, we follow \citet{scranton02} and form the scaled product of elements in each row (or column) $i$ of the correlation matrix: $P(i)=\prod_{r=1}^{N}| {\rm Corr}(\theta_i,\theta_r)|^{1/N}$. This product will be equal to one in the case of perfect correlation, while we find $P\la 0.2$ at all scales of the correlation matrix, indicative of relatively small correlations. Thus, the \lq true\rq\ significance level for the ACF of the hard-spectrum sources is likely to be straddled by the 2- (bootstrap) and 3- (Poisson) $\sigma$ levels, as calculated above.

\begin{table}
 \begin{center}
 \begin{tabular}{llcr}
   \multicolumn{2}{c}{Selection Criteria}                         &  K-S &  ACF\\
   \hline
   S/N$>$3 (B+G)         &     0.5--2 keV           &  0.001  &   $\theta_0=6.3''\pm 3$;\\
                         &                          &         &   $\gamma=2.2\pm 0.2$ \\
   S/N$>$3 (B+G)         &     2--10  keV           &  0.55   &     --   \\
  S/N$>$2 (G)            &     ~~~~~~$''$~~~~~~~;  1$\ge$\hr$>$--0.2   &  0.006   &   $\theta_0=42''^{+7}_{-13}$; \\
                         &                          &         &   $\gamma=3.1_{-0.5}^{+1.1}$  \\
   ~~~~~~~~$''$          &     ~~~~~~$''$~~~~~~~;  1$>$\hr$>$--0.2 &  0.001  &   -- \\
   \hline
 \end{tabular}
 \caption{Basic results of the auto-correlation analysis for various samples. \lq K-S\rq\ refers to the null hypothesis probability of the data and control samples being drawn from the same distribution. Power-law fits (Eq.~\ref{eqn:acf_pl}) to the ACF are listed in the final column, where computed or found to be significant.\label{tab:acfsummary}}
 \end{center}
\end{table}

\section{Discussion}
\label{sec:discussion}

The XMM-LSS survey is the widest, medium-deep, high galactic-latitude X-ray survey carried out by \xmm. 
The survey currently covers a full area of 5.7~\sqd\ and is designed to provide the best constraints on X-ray detected clusters and their evolution out to $z=1$. The wide, contiguous coverage also gives ample opportunity to study the distribution of AGN in this field.

\subsection{Comparison with other works}

We find a positive two-point angular clustering signal with small correlation length in the 0.5--2 keV band, consistent within the errors with previous measurements using similar analysis \citep{vikhlininforman95, basilakos05}.
 We also note that the value of the projected correlation scale that we found is close to the size of the \xmm\ PSF ($\sim 6''$), implying that some level of amplification bias may artificially result in an overestimation of the true clustering scale. The finite PSF can lead to confusion and an effective smoothing of the real source distribution, resulting in a larger correlation length for the density peaks of the observed, smoothed distribution \citep{kaiser84}. We do not expect this bias to have a major effect, though, because the areal density of sources in any spatial resolution element (PSF) is small in  relatively-shallow surveys such as ours (see also \citealt{basilakos05}).

In the hard (2--10 keV) band, 
we do not detect any clustering signal of the projected source distribution on the sky, unlike results by \citet{yang03} and \citet{basilakos04}. The main differences of our survey with respect to the other works are: {\sl i)} the analyzed region covers 4.2~\sqd\ which is at least a factor of two larger than that of \citet{basilakos04} and $\sim$10 times larger than \citet{yang03}; {\sl ii)} this coverage is contiguous (or at least pseudo-contiguous, covering 83 per cent of the area of a $2\times 2.5$~\sqd\ rectangle); {\sl iii)} in terms of flux, while we probe sources slightly deeper than the nominal threshold hard band flux level of $\sim 10^{-14}$ erg s$^{-1}$ cm$^{-2}$ of \citet{basilakos04}, the \c\ survey of \citet{yang03} probes down to $3\times 10^{-15}$ erg s$^{-1}$ cm$^{-2}$, approximately 2--3 times fainter than us\footnote{Both quoted limits are in the 2--8 keV band. The corresponding flux limit in 2--10 keV should be $\approx 20$ per cent brighter, assuming a photon-index $\Gamma=1.7$; the difference in band definitions is not a dominant source of discrepancy.}.

Are we then simply seeing a \lq truer\rq\ picture of the distribution of AGN in X-rays as sky coverage is increased and improved? Or is the XMM-LSS area special in some way? We first note that \lognlogs\ of the XMM-LSS field is slightly lower than other results published in the literature at fluxes brighter than $\sim 2\times 10^{-14}$ erg s$^{-1}$ cm$^{-2}$ in both bands. The field for the XMM-LSS survey was explicitly defined so as to avoid previously-known, bright X-ray sources and the deficit seen could simply reflect cosmic variance in the X-ray sky. Indeed, cosmic variance is known to cause uncertainties in the normalization of the cosmic X-ray background level of $\sim$30 per cent above 2 keV \citep{cowie02, barcons00}. Fluctuations in the {\sl HEAO 1 A-2} X-ray background map on scales of a few degrees have been observed \citep{boughn02, fabianbarcons92} and it may be plausible that our field sits in a comparatively large \lq void\rq\ of large scale structure. The fraction the hard X-ray background resolved out in our field is, however, consistent with that found by other surveys (\S~\ref{sec:xrb}). \citet{yang03} also noted that cosmic variance manifests as \lq voids\rq, but on scales much less than a sq.~deg. 

Due to the sharper \c\ PSF and the deeper flux limit of their survey, the positive result of \citet{yang03} may be understood in terms of their much better sensitivity to detect obscured AGN and any associated clustering signal. Not enough details are available for an assessment of any other differences (such as pipeline selection effects, differences in ACF fitting etc.) with respect to \citet{basilakos04}. But a crude comparison shows a definite deficit of source pairs at small separations in our field. Using their Fig.~1 and Eq.~1, we infer that their fields contain $\sim 9$ source pairs with a separation of between 50 and 60~arcsec over their area of $\approx 2$~\sqd\ at a similar (or slightly higher) flux limit, while we detect exactly 9 such pairs over our full area of 4.2~\sqd. Another difference is that we use the average \lognlogs\ of the XMM-LSS field itself to generate the fluxes of the random catalogues, while they use the distribution from another field (that of \citealt{baldi02}), though their data is consistent with the distribution used (at least at fluxes brighter than $5\times 10^{-14}$ erg s$^{-1}$ cm$^{-2}$; see their Fig~1). We also note that \citet{basilakos04} do not include shot noise (\S~\ref{sec:randomcatalogues}) in the measurement of their random fluxes, but we found that switching off this extra Poisson noise during the generation of our random fluxes had only minor effects on the determination of the ACF.
Finally, any bins with a negative $\omega$ contribution at small scales (which often occur in sparse, weakly-clustered samples) are neglected by them. 
Full source characterization and follow-up in our field should help to discern the correct cause of the observed differences.

\subsection{Selection and correlation of obscured AGN}
\label{sec:hardsources_discussion}

We have made predictions for the expected distribution of \hr\ that would be observed for AGN spread in redshift and affected by different columns of intrinsic obscuring gas. We assumed that our sample follows a luminosity function (XLF) and obscuring column density (\nh) distribution as calculated in recent work (\citealt{ueda03}), and folded the number counts predicted for a limiting flux $S_{\lim}^{2-10}=8\times 10^{-15}$ erg s$^{-1}$ cm$^{-2}$ through the EPIC response function (for simplicity, we used the pn on-axis response only). 
As input templates for this calculation, we used power-law X-ray spectra with a fixed, intrinsic photon-index of $\Gamma=1.9$ \citep[e.g., ][]{mateos05} and modelled photoelectric absorption and Compton reflection \citep{g03}, but did not include Compton-thick sources ($N_{\rm H}>10^{24}$ cm$^{-2}$). Obscured AGN are defined as those with $N_{\rm H}>10^{22}$ cm$^{-2}$. The first plot in Fig.~\ref{fig:hr_expectation} shows that the dominant contributors to the sample of objects with \hr$>$--0.2 (hard-spectrum sources) will be obscured AGN. Intrinsic photon-index variations may push some {\em un}obscured AGN into the high \hr\ regime. Moreover, obscured AGN at high redshift are likely to have low \hr\ values (due to positive $k$-correction into the soft band), so this threshold definitely does not select all obscured AGN uniquely; but given the relatively-shallow depth of our survey, obscured AGN will still dominate this sample. Also note that our observed sample of hardness ratios in Fig.~\ref{fig:hrcts} shows an increase of \hr\ with decreasing flux, as is typically associated with obscured AGN in which the soft-band flux is depleted due to photo-electric absorption in the torus.

Over the deepest region of our survey, we detect marginal, but positive correlation at the $\sim 2-3\sigma$ level for the sample of hard-spectrum sources, at angular separations below $\sim 100$ arcsec. Interestingly, the values of $\omega$ at these separations match well those inferred by \citet{basilakos04} and \citet{yang03} for the full hard band, which might suggest cosmic variance of obscured AGN specifically. Since obscured AGN begin to dominate the source counts only at faint fluxes, we had to include sources with a less stringent detection significance (S/N$>$2) for this analysis. 
We can be confident that most of these are not spurious because 
a large fraction (85 per cent) also have associated 0.5--2 keV counterparts (see Fig.~\ref{fig:hrcts}).
The most suspect detections are the hardest, faint ones with no 0.5--2 keV counterparts, but the correlation signal is not adversely affected by removal of these sources (\S~\ref{sec:hardsources}). Finally, we note that if this sample were dominated by spurious fluctuations, any intrinsic clustering signal would have been diminished, rather than enhanced as we observe, due to their Poisson nature.

\subsubsection{Inversion of the Limber equation}

\begin{figure*}
  \begin{center}
    \includegraphics[angle=90,width=8.6cm]{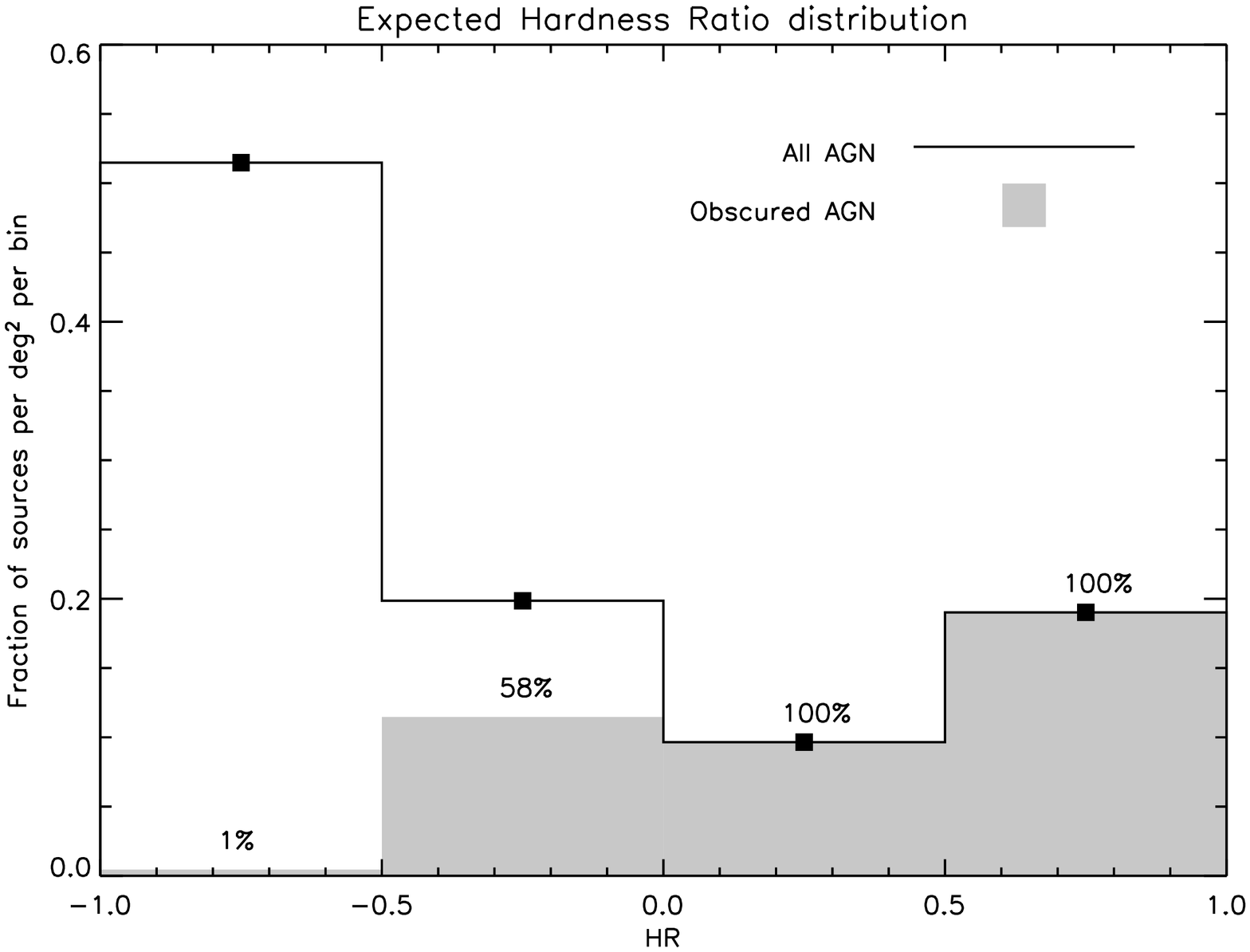}
    \includegraphics[angle=90,width=8.6cm]{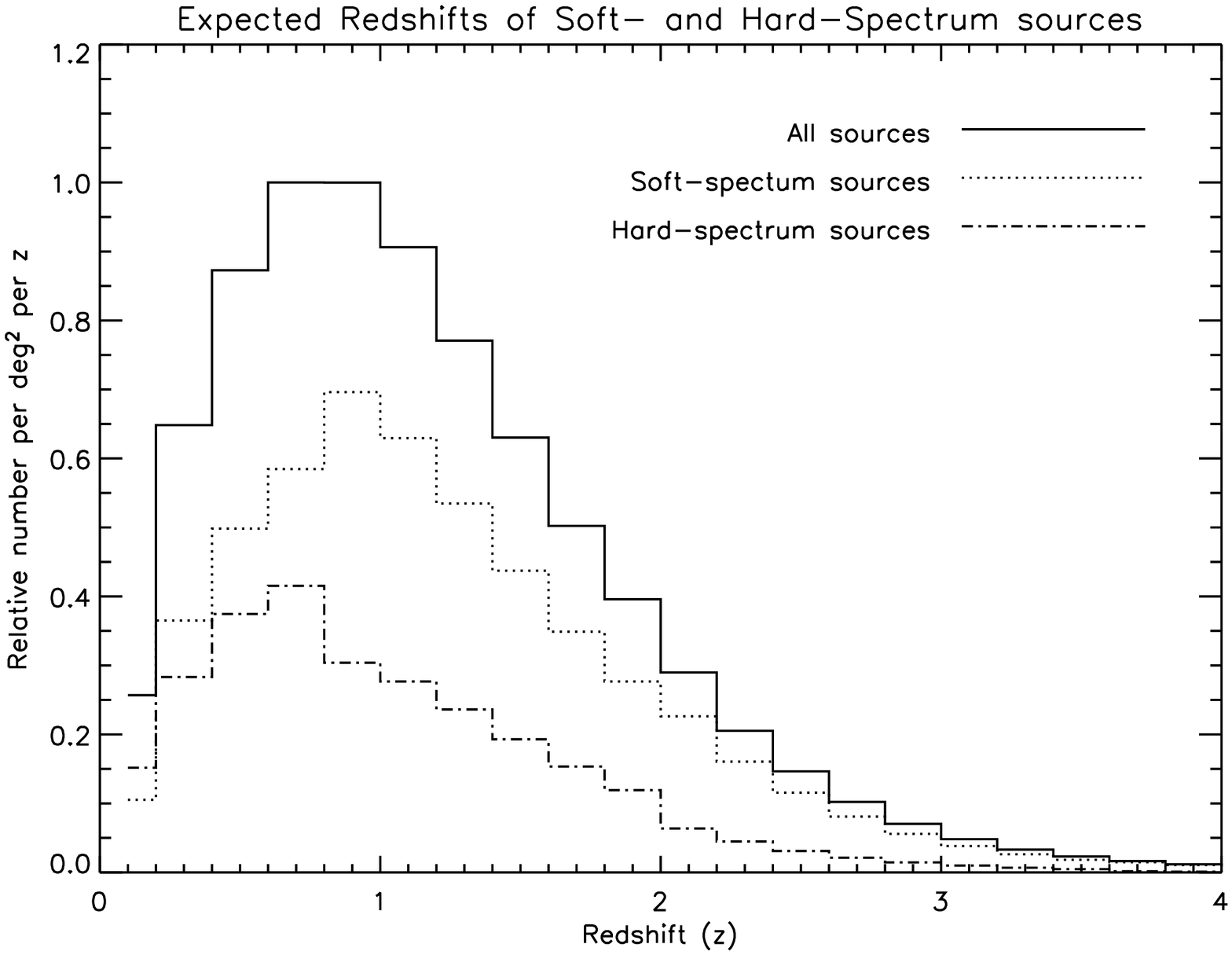}
  \caption{Predictions of the distribution and classification of AGN as soft- and hard-spectrum sources in the XMM-LSS. ({\sl Left}) The plot shows the expected hardness ratio histogram of AGN with fluxes$> 8\times 10^{-15}$ erg s$^{-1}$ cm$^{-2}$ in the 2--10 keV band, assuming the XLF and \nh\ distribution of \citet{ueda03}. The EPIC-pn camera response with the Thin filter is assumed. The fraction of obscured AGN ($N_{\rm H}> 10^{22}$ cm$^{-2}$) is shaded in grey and their relative percentage of all sources in each bin is labelled. While most ($\gtsim 50$ per cent) of the detected sources have very low values of \hr\ (and lie in the first bin), the plot shows that at high \hr\ values, obscured AGN completely dominate. ({\sl Right}) The plot shows the expected redshift distribution of AGN classified as soft-spectrum and hard-spectrum sources above the same flux limit, assuming the luminosity-dependent density evolution model of \citeauthor{ueda03} Hard-spectrum sources peak at $z \sim 0.7$.\label{fig:hr_expectation}} 
  \end{center}
\end{figure*}

An estimate of the real space correlation function can be obtained by inversion of the Limber equation that connects the true space correlation scale $r_0$ with the projected scale $\theta_0$ (see, e.g., \citealt{peebles80, wilman03}). For this, an estimate of the source redshift distribution and a model for the clustering evolution is required. 
We assume that the AGN in our survey follow the expected redshift distribution based on the \citet{ueda03} model as described above. Predictions can be made separately for sources that would be classified as either soft-spectrum, or hard-spectrum AGN, based on their count ratios, irrespective of whether they are intrinsically obscured or unobscured. This redshift distribution is shown in Fig.~\ref{fig:hr_expectation}. Hard-spectrum sources peak at $z=0.7$, with a detectable tail extending out to $z\sim2$ at least. Though the power-law model of \S~\ref{sec:hardsources} is not a good model to fit, assuming the best-fit parameters found therein, and inverting the Limber equation, 
we find $r_0=6(\pm 3) h^{-1}$ Mpc [$H_0=100 h$ km s$^{-1}$ Mpc$^{-1}$] by assuming co-moving clustering evolution and the redshift distribution of hard-spectrum sources in Fig.~\ref{fig:hr_expectation}. This scale is similar to that of local, optically-selected galaxies \citep{davispeebles83} or that of optically-selected QSOs at $z\sim 1$ \citep{croom01}. On the other hand, it is also consistent at the $\sim 2 \sigma$ level with the stronger clustering of extremely-red objects and powerful radio galaxies (e.g., \citealt{rottgering03} and references therein). Given the weakness of the contraints above, however, we defer further discussion on this until source follow-up is complete.

\subsection{Implications}

The typical luminosities of AGN probed in medium-deep surveys such as the XMM-LSS will be $L_{\rm X-ray} \gtsim 10^{43.5}$ erg~s$^{-1}$ \citep[e.g., ][]{g04}. A small correlation length is then consistent with X-ray selected Seyferts being unbiased tracers of structure. This is also in qualitative agreement with studies such as that of \citet{waskett05}, who found no difference in the environments of AGN as compared to those of inactive galaxies. \citet{wake04} also arrived at a similar conclusion from a large sample of optically-selected AGN.

On the other hand, if the auto-correlation of hard-spectrum X-ray sources is indeed larger, it would suggest that obscured AGN are preferentially associated with higher density peaks in the underlying matter distribution; the large amount of gas and dust present in the environment not only triggers AGN activity, but also hides the AGN itself. Galaxy mergers could provide the gas necessary to achieve both (e.g., \citealt{hopkins05}), as is the case for the population of Ultra Luminous Infra Red Galaxies. In fact, assuming that AGN found in X-rays are correlated with powerful-infrared, obscured starbursts known to peak at $z\sim 0.7$ (e.g., \citealt{ce01}), models can be constructed to explain the X-ray background spectrum as well as X-ray source counts \citep{franceschini02, g03}. The mechanism that could drive the gas and dust from the large-scale environment to the scales of galactic nuclei remains unclear, however. Assuming a median redshift of 0.7, angular separations of 50--100~arcsec (the first two bins of the ACF in Fig.~\ref{fig:acf_hardsources}) correspond to projected physical separations of $\sim 350 - 700$ kpc, too large for these systems to be bound mergers in advanced stages. Associations with filaments and groupings in the large scale matter distribution is a possibility. Interactions with any enhanced density of minor galaxies also associated with the large scale structure could provide the necessary torques to trigger gas inflows towards the nucleus, via bars or bound mergers, for instance \citep{shlosman89, derobertis98}.

In contrast, more recent X-ray luminosity function determinations and background synthesis models (over shallower areas of sky at the relevant flux limits, but with extensive redshift coverage) do not require separate evolutionary and/or formation scenarios for obscured and unobscured AGN \citep[e.g., ][]{treister04, ueda03}: these would predict no difference in the comparative correlation of obscured and unobscured AGN. Given the low overall significance of our detected correlation signal for hard-spectrum sources, we can certainly not rule out this possibility. Indeed, in a recent spatial clustering analysis, \citet{yang06} found no difference in the correlation properties of hard- and soft-spectrum AGN (see also \citealt{gilli05}), in contrast to their previous results in angular coordinates for the same sample. This difference might be the result of dissimilar redshift distributions of the two classes of sources; whatever the reason, it underscores the need for complete studies over larger areas of sky. 

Other on-going works which will be able to measure angular correlations include the AXIS \citep{carrera06} and the COSMOS surveys \citep{hasinger06}. The forthcoming expansion of the XMM-LSS survey itelf, along with its deep, multi-wavelength follow-up will provide good constraints on the spatial distribution and clustering of AGN as well as clusters. With an area of 10 deg$^2$, we expect to find at least 1000 AGN above a 2--10 keV flux limit of $8 \times 10^{-15}$ erg s$^{-1}$ cm$^{-2}$ (corresponding to our S/N$>$3 criterion). Additionally, the numbers of obscured AGN detected should double compared to our present sample. The uncertainties on clustering statistics will decrease by a further factor of $\sim$2, giving a proper determination of the slope and scale length and a much better account of the cosmic variance. 

\section{Acknowledgments}

PG thanks the European Southern Observatory Fellowship programme, and the X-ray group at the Institute of Astronomy in Cambridge for their hospitality and computer support during the final stages of this work. We thank the anonymous referee for suggestions that improved the robustness of our procedures. X-ray processing and extensive simulations were performed at the Centre de Calcul de l'IN2P3 in Lyon. The Li\`{e}ge team acknowledges support from PRODEX (XMM). Part of the research was also performed in the framework of the IUAP P5/36 project, supported by the OSTC Belgian Federal services.  LC and DM acknowledge financial contribution from contract ASI-INAF I/023/05/0.

\bibliographystyle{aa}                       
\bibliography{gandhietal_xmmlss.bbl}

\begin{thebibliography}{74}
\expandafter\ifx\csname natexlab\endcsname\relax\def\natexlab#1{#1}\fi

\bibitem[{{Andreon} {et~al.}(2005){Andreon}, {Valtchanov}, {Jones}, {Altieri},
  {Bremer}, {Willis}, {Pierre}, \& {Quintana}}]{andreon05}
{Andreon}, S., {Valtchanov}, I., {Jones}, L.~R., {et~al.} 2005, \mnras, 359,
  1250

\bibitem[{{Arnaud}(1996)}]{xspec}
{Arnaud}, K.~A. 1996, in ASP Conf. Ser. 101: Astronomical Data Analysis
  Software and Systems V, eds. George H. Jacoby and Jeannette Barnes, Vol.~5,
  17--+

\bibitem[{{Baldi} {et~al.}(2002){Baldi}, {Molendi}, {Comastri}, {Fiore},
  {Matt}, \& {Vignali}}]{baldi02}
{Baldi}, A., {Molendi}, S., {Comastri}, A., {et~al.} 2002, \apj, 564, 190

\bibitem[{{Barcons} {et~al.}(2000){Barcons}, {Mateos}, \&
  {Ceballos}}]{barcons00}
{Barcons}, X., {Mateos}, S., \& {Ceballos}, M.~T. 2000, \mnras, 316, L13

\bibitem[{{Barrow} {et~al.}(1984){Barrow}, {Bhavsar}, \& {Sonoda}}]{bootstrap}
{Barrow}, J.~D., {Bhavsar}, S.~P., \& {Sonoda}, D.~H. 1984, \mnras, 210, 19P

\bibitem[{{Basilakos} {et~al.}(2004){Basilakos}, {Georgakakis}, {Plionis}, \&
  {Georgantopoulos}}]{basilakos04}
{Basilakos}, S., {Georgakakis}, A., {Plionis}, M., \& {Georgantopoulos}, I.
  2004, \apjl, 607, L79

\bibitem[{{Basilakos} {et~al.}(2005){Basilakos}, {Plionis}, {Georgakakis}, \&
  {Georgantopoulos}}]{basilakos05}
{Basilakos}, S., {Plionis}, M., {Georgakakis}, A., \& {Georgantopoulos}, I.
  2005, \mnras, 356, 183

\bibitem[{{Bertin} \& {Arnouts}(1996)}]{sextractor}
{Bertin}, E. \& {Arnouts}, S. 1996, \aaps, 117, 393

\bibitem[{{Boughn} {et~al.}(2002){Boughn}, {Crittenden}, \&
  {Koehrsen}}]{boughn02}
{Boughn}, S.~P., {Crittenden}, R.~G., \& {Koehrsen}, G.~P. 2002, \apj, 580, 672

\bibitem[{{Brandt} {et~al.}(2004){Brandt}, {Alexander}, {Bauer}, \&
  {Vignali}}]{brandt_chile}
{Brandt}, W.~N., {Alexander}, D.~M., {Bauer}, F.~E., \& {Vignali}, C. 2004,
  Physics of Active Galactic Nuclei at All Scales, eds. Alloin D., Johnson R.,
  Lira P. (Springer-Verlag, Berlin), astro-ph/0403646

\bibitem[{{Brandt} \& {Hasinger}(2005)}]{brandthasinger05}
{Brandt}, W.~N. \& {Hasinger}, G. 2005, \araa, 43, 827

\bibitem[{{Carrera} {et~al.}(1998){Carrera}, {Barcons}, {Fabian}, {Hasinger},
  {Mason}, {McMahon}, {Mittaz}, \& {Page}}]{carrera98}
{Carrera}, F.~J., {Barcons}, X., {Fabian}, A.~C., {et~al.} 1998, \mnras, 299,
  229

\bibitem[{{Carrera} {et~al.}(2006){Carrera}, {Ebrero}, {Mateos}, {Ceballos},
  {Corral}, {Mason}, {McMahon}, {Mittaz}, \& {Page}}]{carrera06}
{Carrera}, F.~J., {Ebrero}, S., {Mateos}, S., {et~al.} 2006, in preparation

\bibitem[{{Chary} \& {Elbaz}(2001)}]{ce01}
{Chary}, R. \& {Elbaz}, D. 2001, \apj, 556, 562

\bibitem[{{Chiappetti} {et~al.}(2005){Chiappetti}, {Tajer}, {Trinchieri},
  {Maccagni}, {Maraschi}, {Paioro}, {Pierre}, {Surdej}, {Garcet}, {Gosset}, {Le
  F{\` e}vre}, {Bertin}, {McCracken}, {Mellier}, {Foucaud}, {Radovich},
  {Ripepi}, \& {Arnaboldi}}]{chiappetti05}
{Chiappetti}, L., {Tajer}, M., {Trinchieri}, G., {et~al.} 2005, \aap, 439, 413

\bibitem[{{Cohen} {et~al.}(2003){Cohen}, {R{\" o}ttgering}, {Kassim}, {Cotton},
  {Perley}, {Wilman}, {Best}, {Pierre}, {Birkinshaw}, {Bremer}, \&
  {Zanichelli}}]{cohen03}
{Cohen}, A.~S., {R{\" o}ttgering}, H.~J.~A., {Kassim}, N.~E., {et~al.} 2003,
  \apj, 591, 640

\bibitem[{{Cowie} {et~al.}(2002){Cowie}, {Garmire}, {Bautz}, {Barger},
  {Brandt}, \& {Hornschemeier}}]{cowie02}
{Cowie}, L.~L., {Garmire}, G.~P., {Bautz}, M.~W., {et~al.} 2002, \apjl, 566, L5

\bibitem[{{Crawford} {et~al.}(2002){Crawford}, {Gandhi}, {Fabian}, {Wilman},
  {Johnstone}, {Barger}, \& {Cowie}}]{c02}
{Crawford}, C.~S., {Gandhi}, P., {Fabian}, A.~C., {et~al.} 2002, \mnras, 333,
  809

\bibitem[{{Croom} {et~al.}(2001){Croom}, {Shanks}, {Boyle}, {Smith}, {Miller},
  {Loaring}, \& {Hoyle}}]{croom01}
{Croom}, S.~M., {Shanks}, T., {Boyle}, B.~J., {et~al.} 2001, \mnras, 325, 483

\bibitem[{{Davis} \& {Peebles}(1983)}]{davispeebles83}
{Davis}, M. \& {Peebles}, P.~J.~E. 1983, \apj, 267, 465

\bibitem[{{De Luca} \& {Molendi}(2004)}]{delucamolendi04}
{De Luca}, A. \& {Molendi}, S. 2004, \aap, 419, 837

\bibitem[{{de Robertis} {et~al.}(1998){de Robertis}, {Yee}, \&
  {Hayhoe}}]{derobertis98}
{de Robertis}, M.~M., {Yee}, H.~K.~C., \& {Hayhoe}, K. 1998, \apj, 496, 93

\bibitem[{{Efstathiou} {et~al.}(1991){Efstathiou}, {Bernstein}, {Tyson},
  {Katz}, \& {Guhathakurta}}]{efstathiou91}
{Efstathiou}, G., {Bernstein}, G., {Tyson}, J.~A., {Katz}, N., \&
  {Guhathakurta}, P. 1991, \apjl, 380, L47

\bibitem[{{Fabian} \& {Barcons}(1992)}]{fabianbarcons92}
{Fabian}, A.~C. \& {Barcons}, X. 1992, \araa, 30, 429

\bibitem[{{Fisher} {et~al.}(1994){Fisher}, {Davis}, {Strauss}, {Yahil}, \&
  {Huchra}}]{fisher94}
{Fisher}, K.~B., {Davis}, M., {Strauss}, M.~A., {Yahil}, A., \& {Huchra}, J.
  1994, \mnras, 266, 50

\bibitem[{{Franceschini} {et~al.}(2002){Franceschini}, {Braito}, \&
  {Fadda}}]{franceschini02}
{Franceschini}, A., {Braito}, V., \& {Fadda}, D. 2002, \mnras, 335, L51

\bibitem[{{Gandhi} {et~al.}(2004){Gandhi}, {Crawford}, {Fabian}, \&
  {Johnstone}}]{g04}
{Gandhi}, P., {Crawford}, C.~S., {Fabian}, A.~C., \& {Johnstone}, R.~M. 2004,
  \mnras, 348, 529

\bibitem[{{Gandhi} \& {Fabian}(2003)}]{g03}
{Gandhi}, P. \& {Fabian}, A.~C. 2003, \mnras, 339, 1095

\bibitem[{{Gehrels}(1986)}]{gehrels86}
{Gehrels}, N. 1986, \apj, 303, 336

\bibitem[{{Gilli} {et~al.}(2005){Gilli}, {Daddi}, {Zamorani}, {Tozzi},
  {Borgani}, {Bergeron}, {Giacconi}, {Hasinger}, {Mainieri}, {Norman},
  {Rosati}, {Szokoly}, \& {Zheng}}]{gilli05}
{Gilli}, R., {Daddi}, E., {Zamorani}, G., {et~al.} 2005, \aap, 430, 811

\bibitem[{{Hamilton}(1993)}]{hamilton93}
{Hamilton}, A.~J.~S. 1993, \apj, 417, 19

\bibitem[{{Hasinger} {et~al.}(2006){Hasinger}, {Cappelluti,}, {Brunner},
  {Barcons}, {Bergeron}, {Brunner}, {Dadina}, {Dennerl}, {Ferrando},
  {Finoguenov}, {Griffiths}, {Hashimoto}, {Jansen}, {Lumb}, {Mason}, {Mateos},
  {McMahon}, {Miyaji}, {Paerels}, {Page}, {Ptak}, {Sasseen}, {Schartel},
  {Szokoly}, {Tr{\" u}mper}, {Turner}, {Warwick}, \& {Watson}}]{hasinger06}
{Hasinger}, G., {Cappelluti,}, N., {Brunner}, H., {et~al.} 2006, \apjs\
  submitted

\bibitem[{{Hickox} \& {Markevitch}(2006)}]{hickoxmarkevitch06}
{Hickox}, R.~C. \& {Markevitch}, M. 2006, ApJ submitted; arXiv:astro-ph/0512542

\bibitem[{{Hopkins} {et~al.}(2005){Hopkins}, {Hernquist}, {Cox}, {Di Matteo},
  {Martini}, {Robertson}, \& {Springel}}]{hopkins05}
{Hopkins}, P.~F., {Hernquist}, L., {Cox}, T.~J., {et~al.} 2005, \apj, 630, 705

\bibitem[{{Joyce} {et~al.}(1999){Joyce}, {Montuori}, \&
  {Labini}}]{clustering_fractals}
{Joyce}, M., {Montuori}, M., \& {Labini}, F.~S. 1999, \apjl, 514, L5

\bibitem[{{Kaiser}(1984)}]{kaiser84}
{Kaiser}, N. 1984, \apjl, 284, L9

\bibitem[{{Le F{\` e}vre} {et~al.}(2004){Le F{\` e}vre}, {Mellier},
  {McCracken}, {Foucaud}, {Gwyn}, {Radovich}, {Dantel-Fort}, {Bertin},
  {Moreau}, {Cuillandre}, {Pierre}, {Le Brun}, {Mazure}, \&
  {Tresse}}]{lefevre04}
{Le F{\` e}vre}, O., {Mellier}, Y., {McCracken}, H.~J., {et~al.} 2004, \aap,
  417, 839

\bibitem[{{Lonsdale} {et~al.}(2003){Lonsdale}, {Smith}, {Rowan-Robinson},
  {Surace}, {Shupe}, {Xu}, {Oliver}, {Padgett}, {Fang}, {Conrow},
  {Franceschini}, {Gautier}, {Griffin}, {Hacking}, {Masci}, {Morrison},
  {O'Linger}, {Owen}, {P{\' e}rez-Fournon}, {Pierre}, {Puetter}, {Stacey},
  {Castro}, {Del Carmen Polletta}, {Farrah}, {Jarrett}, {Frayer}, {Siana},
  {Babbedge}, {Dye}, {Fox}, {Gonzalez-Solares}, {Salaman}, {Berta}, {Condon},
  {Dole}, \& {Serjeant}}]{lonsdale03}
{Lonsdale}, C.~J., {Smith}, H.~E., {Rowan-Robinson}, M., {et~al.} 2003, \pasp,
  115, 897

\bibitem[{{Maiolino} \& {Rieke}(1995)}]{maiolinorieke95}
{Maiolino}, R. \& {Rieke}, G.~H. 1995, \apj, 454, 95+

\bibitem[{{Manners} {et~al.}(2003){Manners}, {Johnson}, {Almaini}, {Willott},
  {Gonzalez-Solares}, {Lawrence}, {Mann}, {Perez-Fournon}, {Dunlop}, {McMahon},
  {Oliver}, {Rowan-Robinson}, \& {Serjeant}}]{manners03}
{Manners}, J.~C., {Johnson}, O., {Almaini}, O., {et~al.} 2003, \mnras, 343, 293

\bibitem[{{Mateos} {et~al.}(2005){Mateos}, {Barcons}, {Carrera}, {Ceballos},
  {Caccianiga}, {Lamer}, {Maccacaro}, {Page}, {Schwope}, \&
  {Watson}}]{mateos05}
{Mateos}, S., {Barcons}, X., {Carrera}, F.~J., {et~al.} 2005, \aap, 433, 855

\bibitem[{{Matt} {et~al.}(2000){Matt}, {Fabian}, {Guainazzi}, {Iwasawa},
  {Bassani}, \& {Malaguti}}]{matt00}
{Matt}, G., {Fabian}, A.~C., {Guainazzi}, M., {et~al.} 2000, \mnras, 318, 173

\bibitem[{{Mullis} {et~al.}(2004){Mullis}, {Henry}, {Gioia}, {B{\"o}hringer},
  {Briel}, {Voges}, \& {Huchra}}]{mullis04}
{Mullis}, C.~R., {Henry}, J.~P., {Gioia}, I.~M., {et~al.} 2004, \apj, 617, 192

\bibitem[{{Mushotzky} {et~al.}(2000){Mushotzky}, {Cowie}, {Barger}, \&
  {Arnaud}}]{mushotzky00}
{Mushotzky}, R.~F., {Cowie}, L.~L., {Barger}, A.~J., \& {Arnaud}, K.~A. 2000,
  \nat, 404, 459

\bibitem[{{Nandra} {et~al.}(2004){Nandra}, {Georgantopoulos}, {Brotherton}, \&
  {Papadakis}}]{nandra04}
{Nandra}, K., {Georgantopoulos}, I., {Brotherton}, M., \& {Papadakis}, I.~E.
  2004, \mnras, 347, L41

\bibitem[{{Pacaud} {et~al.}(2006){Pacaud}, {Pierre}, {Refregier}, {Gueguen},
  {Starck}, {Valtchanov}, {Read}, {Altieri}, {Chiappetti}, {Gandhi}, {Garcet},
  {Gosset}, {Ponman}, \& {Surdej}}]{pacaud06}
{Pacaud}, F., {Pierre}, M., {Refregier}, A., {et~al.} 2006, MNRAS accepted

\bibitem[{{Padovani} {et~al.}(2004){Padovani}, {Allen}, {Rosati}, \&
  {Walton}}]{padovani04}
{Padovani}, P., {Allen}, M.~G., {Rosati}, P., \& {Walton}, N.~A. 2004, \aap,
  424, 545

\bibitem[{{Peebles}(1980)}]{peebles80}
{Peebles}, P.~J.~E. 1980, {The large-scale structure of the universe}
  (Princeton University Press,)

\bibitem[{{Pierre} {et~al.}(2006){Pierre}, {Pacaud}, {Duc}, {Willis},
  {Andreon}, {Valtchanov}, {Altieri}, {Galaz}, {Gueguen}, {Le F\'{e}vre}, {Le
  F\'{e}vre}, {Ponman}, {Sprimont}, {Surdej}, \& {Read}}]{pierre06}
{Pierre}, M., {Pacaud}, F., {Duc}, P.-A., {et~al.} 2006, MNRAS accepted

\bibitem[{{Pierre} {et~al.}(2004){Pierre}, {Valtchanov}, {Santos}, {Altieri},
  {Andreon}, {Bolzonella}, {Bremer}, {Disseau}, {Gandhi}, {Jean}, {Read},
  {Refregier}, {Willis}, {Adami}, {Alloin}, {Birkinshaw}, {Chiappetti},
  {Cohen}, {Detal}, {Duc}, {Gosset}, {Jones}, {Fevre}, {Maccagni}, {McBreen},
  {McCracken}, {Mellier}, {Ponman}, {Quintana}, {Rottgering}, {Smette},
  {Surdej}, {Vigroux}, {Bohringer}, {Hjorth}, {Lonsdale}, \&
  {White}}]{pierre_lss}
{Pierre}, M., {Valtchanov}, I., {Santos}, S.~D., {et~al.} 2004, J. Cosm.
  Astropart. Phys., 9, 11

\bibitem[{{Read} \& {Ponman}(2003)}]{readponman03}
{Read}, A.~M. \& {Ponman}, T.~J. 2003, \aap, 409, 395

\bibitem[{{Refregier} {et~al.}(2002){Refregier}, {Valtchanov}, \&
  {Pierre}}]{refregier02}
{Refregier}, A., {Valtchanov}, I., \& {Pierre}, M. 2002, \aap, 390, 1

\bibitem[{{Rosati} {et~al.}(2002){Rosati}, {Borgani}, \&
  {Norman}}]{rosati02_araa}
{Rosati}, P., {Borgani}, S., \& {Norman}, C. 2002, \araa, 40, 539

\bibitem[{{R{\"o}ttgering} {et~al.}(2003){R{\"o}ttgering}, {Daddi}, {Overzier},
  \& {Wilman}}]{rottgering03}
{R{\"o}ttgering}, H., {Daddi}, E., {Overzier}, R., \& {Wilman}, R. 2003, New
  Astronomy Review, 47, 309

\bibitem[{{Schmidt} {et~al.}(1998){Schmidt}, {Hasinger}, {Gunn}, {Schneider},
  {Burg}, {Giacconi}, {Lehmann}, {MacKenty}, {Trumper}, \&
  {Zamorani}}]{schmidt98}
{Schmidt}, M., {Hasinger}, G., {Gunn}, J., {et~al.} 1998, \aap, 329, 495

\bibitem[{{Scranton} {et~al.}(2002){Scranton}, {Johnston}, {Dodelson},
  {Frieman}, {Connolly}, {Eisenstein}, {Gunn}, {Hui}, {Jain}, {Kent},
  {Loveday}, {Narayanan}, {Nichol}, {O'Connell}, {Scoccimarro}, {Sheth},
  {Stebbins}, {Strauss}, {Szalay}, {Szapudi}, {Tegmark}, {Vogeley}, {Zehavi},
  {Annis}, {Bahcall}, {Brinkman}, {Csabai}, {Hindsley}, {Ivezic}, {Kim},
  {Knapp}, {Lamb}, {Lee}, {Lupton}, {McKay}, {Munn}, {Peoples}, {Pier},
  {Richards}, {Rockosi}, {Schlegel}, {Schneider}, {Stoughton}, {Tucker},
  {Yanny}, \& {York}}]{scranton02}
{Scranton}, R., {Johnston}, D., {Dodelson}, S., {et~al.} 2002, \apj, 579, 48

\bibitem[{{Setti} \& {Woltjer}(1989)}]{settiwoltjer89}
{Setti}, G. \& {Woltjer}, L. 1989, \aap, 224, L21

\bibitem[{{Shlosman} {et~al.}(1989){Shlosman}, {Frank}, \&
  {Begelman}}]{shlosman89}
{Shlosman}, I., {Frank}, J., \& {Begelman}, M.~C. 1989, \nat, 338, 45

\bibitem[{{Spergel} {et~al.}(2003){Spergel}, {Verde}, {Peiris}, {Komatsu},
  {Nolta}, {Bennett}, {Halpern}, {Hinshaw}, {Jarosik}, {Kogut}, {Limon},
  {Meyer}, {Page}, {Tucker}, {Weiland}, {Wollack}, \& {Wright}}]{wmap}
{Spergel}, D.~N., {Verde}, L., {Peiris}, H.~V., {et~al.} 2003, \apjs, 148, 175

\bibitem[{{Szokoly} {et~al.}(2004){Szokoly}, {Bergeron}, {Hasinger}, {Lehmann},
  {Kewley}, {Mainieri}, {Nonino}, {Rosati}, {Giacconi}, {Gilli}, {Gilmozzi},
  {Norman}, {Romaniello}, {Schreier}, {Tozzi}, {Wang}, {Zheng}, \&
  {Zirm}}]{szokoly04}
{Szokoly}, G.~P., {Bergeron}, J., {Hasinger}, G., {et~al.} 2004, \apjs, 155,
  271

\bibitem[{{Treister} {et~al.}(2004){Treister}, {Urry}, {Chatzichristou},
  {Bauer}, {Alexander}, {Koekemoer}, {Van Duyne}, {Brandt}, {Bergeron},
  {Stern}, {Moustakas}, {Chary}, {Conselice}, {Cristiani}, \&
  {Grogin}}]{treister04}
{Treister}, E., {Urry}, C.~M., {Chatzichristou}, E., {et~al.} 2004, \apj, 616,
  123

\bibitem[{{Ueda} {et~al.}(2003){Ueda}, {Akiyama}, {Ohta}, \& {Miyaji}}]{ueda03}
{Ueda}, Y., {Akiyama}, M., {Ohta}, K., \& {Miyaji}, T. 2003, \apj, 598, 886

\bibitem[{{Valtchanov} {et~al.}(2004){Valtchanov}, {Pierre}, {Willis}, {Dos
  Santos}, {Jones}, {Andreon}, {Adami}, {Altieri}, {Bolzonella}, {Bremer},
  {Duc}, {Gosset}, {Jean}, \& {Surdej}}]{valtchanov04}
{Valtchanov}, I., {Pierre}, M., {Willis}, J., {et~al.} 2004, \aap, 423, 75

\bibitem[{{Vikhlinin} \& {Forman}(1995)}]{vikhlininforman95}
{Vikhlinin}, A. \& {Forman}, W. 1995, \apjl, 455, L109+

\bibitem[{{Wake} {et~al.}(2004){Wake}, {Miller}, {Di Matteo}, {Nichol}, {Pope},
  {Szalay}, {Gray}, {Schneider}, \& {York}}]{wake04}
{Wake}, D.~A., {Miller}, C.~J., {Di Matteo}, T., {et~al.} 2004, \apjl, 610, L85

\bibitem[{{Waskett} {et~al.}(2005){Waskett}, {Eales}, {Gear}, {McCracken},
  {Lilly}, \& {Brodwin}}]{waskett05}
{Waskett}, T.~J., {Eales}, S.~A., {Gear}, W.~K., {et~al.} 2005, MNRAS accepted;
  astro-ph/0508102

\bibitem[{{Webster}(1976)}]{psa}
{Webster}, A. 1976, \mnras, 175, 61

\bibitem[{{Willis} {et~al.}(2005{\natexlab{a}}){Willis}, {Pacaud},
  {Valtchanov}, {Pierre}, {Ponman}, {Read}, {Andreon}, {Altieri}, {Quintana},
  {Dos Santos}, {Birkinshaw}, {Bremer}, {Duc}, {Galaz}, {Gosset}, {Jones}, \&
  {Surdej}}]{willis05}
{Willis}, J.~P., {Pacaud}, F., {Valtchanov}, I., {et~al.} 2005{\natexlab{a}},
  \mnras, 363, 675

\bibitem[{{Willis} {et~al.}(2005{\natexlab{b}}){Willis}, {Pacaud},
  {Valtchanov}, {Pierre}, {Ponman}, {Read}, {Andreon}, {Altieri}, {Quintana},
  {Santos}, {Birkinshaw}, {Bremer}, {Duc}, {Galaz}, {Gosset}, {Jones}, \&
  {Surdej}}]{willis05erratum}
{Willis}, J.~P., {Pacaud}, F., {Valtchanov}, I., {et~al.} 2005{\natexlab{b}},
  \mnras, 364, 751

\bibitem[{{Wilman} {et~al.}(2003){Wilman}, {R{\" o}ttgering}, {Overzier}, \&
  {Jarvis}}]{wilman03}
{Wilman}, R.~J., {R{\" o}ttgering}, H.~J.~A., {Overzier}, R.~A., \& {Jarvis},
  M.~J. 2003, \mnras, 339, 695

\bibitem[{{Worsley} {et~al.}(2004){Worsley}, {Fabian}, {Barcons}, {Mateos},
  {Hasinger}, \& {Brunner}}]{worsley04}
{Worsley}, M.~A., {Fabian}, A.~C., {Barcons}, X., {et~al.} 2004, \mnras, 352,
  L28

\bibitem[{{Yang} {et~al.}(2006){Yang}, {Mushotzky}, {Barger}, \&
  {Cowie}}]{yang06}
{Yang}, Y., {Mushotzky}, R.~F., {Barger}, A.~J., \& {Cowie}, L.~L. 2006, ApJ
  accepted, astro-ph/0601634

\bibitem[{{Yang} {et~al.}(2003){Yang}, {Mushotzky}, {Barger}, {Cowie},
  {Sanders}, \& {Steffen}}]{yang03}
{Yang}, Y., {Mushotzky}, R.~F., {Barger}, A.~J., {et~al.} 2003, \apjl, 585, L85

\bibitem[{{Zacharias} {et~al.}(2004){Zacharias}, {Urban}, {Zacharias},
  {Wycoff}, {Hall}, {Monet}, \& {Rafferty}}]{usno}
{Zacharias}, N., {Urban}, S.~E., {Zacharias}, M.~I., {et~al.} 2004, \aj, 127,
  3043

\end{thebibliography}

\label{lastpage}

\end{document}